\documentclass[12pt]{article}
\usepackage{amsmath,amssymb,epsfig,ulem}
\usepackage{graphicx,floatflt,subfigure}
\usepackage{cite}

\makeatletter

\newcommand{\be}{\begin{equation}}
\newcommand{\ee}{\end{equation}}
\newcommand{\ba}{\begin{aligned}}
\newcommand{\ea}{\end{aligned}}

\def\cC{\cal{C}}

\def\adj{\mathop{\mathrm{adj}}\nolimits}

\def\unit{{1\kern-.65ex {\rm l}}}
\def\1{{1\kern-.65ex {\rm l}}}

\newcount\hour \newcount\minute
\hour=\time \divide \hour by 60
\minute=\time
\def\now{%
\ifnum \hour<13
  \ifnum \hour=0 \advance \hour by 12 \number\hour:\else \number\hour:\fi%
     \ifnum \minute<10 0\fi%
     \number\minute%
\ A.M.%
\else \advance \hour by -12 \number\hour:%
  \ifnum \minute<10 0\fi%
  \number\minute%
  \ P.M.%
\fi%
}

\makeatother

\begin{document}

\baselineskip=18pt  
\numberwithin{equation}{section}  
\allowdisplaybreaks  



%
%


\thispagestyle{empty}

\vspace*{-2cm}
\begin{flushright}
{\tt arXiv:0906.4672}\\
CALT-68-2733\\
\end{flushright}

\vspace*{0.8cm}
\begin{center}
 {\LARGE Monodromies, Fluxes, and Compact Three-Generation F-theory GUTs\\}
 \vspace*{1.5cm}
 Joseph Marsano, Natalia Saulina, and Sakura Sch\"afer-Nameki\\
 \vspace*{1.0cm}
{\it California Institute of Technology \\
1200 E California Blvd, Pasadena, CA 91125, USA}\\[1ex]
 {\tt marsano, saulina, ss299  theory.caltech.edu}
  \vspace*{0.8cm}
\end{center}
\vspace*{.5cm}

\noindent
We analyze constraints for embedding local $SU(5)$ F-theory GUTs into consistent compactifications and construct explicit three-generation models based on the geometry of \cite{Marsano:2009ym}. The key tool for studying constraints in this problem when there is an underlying $E_8$ structure is the spectral cover, which encodes all of the symmetries that fix the allowed couplings in the superpotential, as well as  the consistent, supersymmetric $G$-fluxes. 
Imposing phenomenological requirements such as the existence of three generations, top and  bottom Yukawa couplings, good flavor structure and absence of exotics and of a tree-level $\mu$-term, we derive stringent constraints on the allowed spectral covers. The resulting spectral covers are in conflict with the neutrino scenarios that have been studied in local F-theory models unless we allow for the possibility of additional charged fields, perhaps playing the role of gauge messengers, that do not comprise complete GUT multiplets.  Quite remarkably, the existence of additional incomplete GUT multiplets below the GUT scale is necessary for consistency with gauge coupling "unification", as their effect can precisely cancel that of the internal hypercharge flux, which distorts the gauge couplings already at $M_{\rm GUT}$.

\newpage
\setcounter{page}{1} 



\tableofcontents


\section{Introduction and Summary}

Over the past year, it has become increasingly clear that F-theory provides a very promising framework for constructing realistic supersymmetric GUT models in string theory \cite{Donagi:2008ca,Beasley:2008dc}.  Studies of local models have identified, among other things, mechanisms capable of producing doublet-triplet splitting without the usual proton decay problems \cite{Beasley:2008kw,Donagi:2008kj}, natural flavor hierarchies \cite{Heckman:2008qa}, and realistic levels of mixing in the quark \cite{Heckman:2008qa,Heckman:2009de} and neutrino sectors \cite{Bouchard:2009bu,Heckman:2009mn}.  Further, F-theory models admit simple implementations of gauge mediation \cite{Marsano:2008jq,Heckman:2008qt} that come naturally equipped with a mechanism for addressing the $\mu$ and $\mu/B_{\mu}$ problems \cite{Ibe:2007km}.    In this paper, we seek to understand some of the basic constraints that arise when attempting to realize these successes in full F-theory compactifications.  We then aim to construct compact examples that implement as many of these constraints as possible.

The conditions that we demand of our models do not seem that severe at the outset.  Roughly speaking, they amount to requiring
\begin{itemize}
\item GUT-breaking and doublet-triplet splitting via a nontrivial hypercharge flux
\item Realization of the MSSM superpotential
\item Absence of dangerous dimension 4 proton decay operators
\item Absence of a bare $\mu$ term
\item Some of the requisite structure for getting flavor hierarchies
\item No charged exotics
\end{itemize}
We will clarify what we mean by these conditions in section \ref{sec:constraints}.  This list certainly does not comprise all of the features that a successful model must include, but it seems to be a reasonable starting point.

\subsection{Matter Curves and Symmetry Structure}

At first glance, it might seem that these conditions are easy to realize.  This is certainly true in the local picture, wherein one studies the model only in certain patches on the 4-cycle, $S_{GUT}$, on which the GUT degrees of freedom are localized.  There, superpotential couplings are completely determined by the intersection properties of matter curves, over which we have significant control.  Forbidding dimension 4 proton decay operators, for instance, seems pretty easy since we just need to prevent the corresponding matter curves from intersecting in the wrong way.

Already when one tries to obtain several different couplings from a single point of enhanced symmetry, such as the $E_8$ points of \cite{Bouchard:2009bu,Heckman:2009mn}, the situation becomes somewhat more subtle.  When a large number of matter curves come together, $SU(5)$ invariance is not enough to determine which couplings are generated.  Rather, the superpotential depends on how each individual multiplet embeds into the adjoint of $E_8$.  More specifically, each matter multiplet is distinguished by its charges under the $U(1)^4$ Cartan subalgebra of the $SU(5)_{\perp}$, the commutant of $SU(5)$ inside $E_8$.  The allowed couplings, then,  are precisely those that are invariant under this $U(1)^4$.

When one moves to the "semi-local" picture of \cite{Donagi:2009ra}, where one studies ALE fibrations over not just isolated patches but rather the full $S_{\rm GUT}$, this subtlety is, in a sense, extended over the entire surface $S_{GUT}$.  This happens partly because the notion of "distinct" matter curves depends on how the corresponding GUT multiplets are embedded into the adjoint of $E_8$.
When two seemingly different matter curves intersect at a point that does not further increase the rank of the singularity, their wave functions become connected by a nontrivial boundary condition \cite{Tatar:2009jk}.  From the perspective of these wave functions, one should not think of the two curves as distinct but rather as a single curve that happens to "pinch" at a point.  Zero modes are counted as though the two comprise a single curve and wave functions are expected to spread over both components.  The connection to $E_8$ group theory arises by noting that the matter curves of charged fields that embed differently into the $E_8$ adjoint are guaranteed to yield enhancements in rank when they intersect, thereby making them truly distinct.

Geometrically, the 4 $U(1)$ factors that distinguish different types of matter curves and control the superpotential arise from 2-cycles that are resolved as the $E_8$ singularity is unfolded.  In a generic coordinate patch, these factors are all distinct but when the patches are glued together to form the full 4-cycle $S_{\rm GUT}$, they typically undergo a series of monodromies.  This forces us to quotient the theory by the monodromy group, $G$, which will be a subgroup of the Weyl group of $E_8$ that leaves the $SU(5)$ roots invariant.  The quotient removes the distinction between some of the matter curves and gives rise to an intricate symmetry structure in the resulting superpotential that can not always be understood in terms of residual global symmetries alone.  A useful object for studying the monodromy group $G$ is a familiar one from heterotic model-building, namely the so-called spectral cover.  The importance of the spectral cover for F-theory models that do not necessarily admit heterotic duals has been emphasized in several publications in the past year \cite{Donagi:2008kj,Hayashi:2009ge,Donagi:2009ra,Tatar:2009jk}.

\subsection{Constraints and Compact Examples}

In the first part of this paper, we will study our basic list of constraints in this "semi-local" framework.  Perhaps surprisingly, we will find that only three choices for the monodromy group, $G$, are consistent with all of them and, in each case, the embeddings of MSSM matter multiplets into $E_8$ are essentially fixed.  All of these scenarios have exactly one anomalous $U(1)$ gauge symmetry that survives the quotient and it is fixed in each case to be a combination of $U(1)_{B-L}$ and $U(1)_Y$.  An accidental global $U(1)_{PQ}$ symmetry under which $H_u$ and $H_d$ carry identical charge arises at the level of the renormalizable Lagrangian, but can be generically broken by nonrenormalizable operators that originate from physics at or above the GUT scale.

After identifying this limited class of scenarios, we next turn to their realization in full F-theory compactifications.  As a warm-up, we drop the $\mu$ term constraint as this leads to significant simplifications.  We then describe two different constructions of $G$-fluxes in the "semi-local" picture that are capable of engineering three generations of chiral matter.  We are able to realize one of these in an F-theory compactification on an elliptically fibered Calabi-Yau four-fold whose base manifold is the one constructed in \cite{Marsano:2009ym}.  This leads to a 3-generation compact F-theory GUT that realizes all of the constraints enumerated above except for the absence of a bare $\mu$ term.  We describe in general how one can extend the symmetry structure to incorporate this constraint as well.  Carrying this out in explicit examples seems daunting, but we see no obvious obstructions.

So, in the end we are able to construct a relatively simple example of a compact F-theory GUT model with three generations of chiral matter, the MSSM superpotential, a viable mechanism for GUT-breaking, and the absence of problems associated with proton decay.  The limitations of this model, however, go beyond the fact that it exhibits a $\mu$ problem.  For instance, while we realize some of the necessary conditions for getting flavor hierarchies, we do not realize all of them.  Doing so will require, at the very least, a further tuning that we do not investigate here.

\subsection{Evading the Constraints}

More troubling, however, is that the limited class of scenarios that satisfy our general constraints seem to generically have problems with neutrino physics.  This is due to the presence of an exact (perturbative) $U(1)_{B-L}$ and the lack of an exact $U(1)_{PQ}$ symmetry with respect to which $H_u$ and $H_d$ have the same charge and may be an indication that the constraints we impose are too restrictive.  

The most significant of all these constraints, and the only one that it seems reasonable to relax, is the precise manner by which we ensure that all exotics are removed from the spectrum.  In particular, we require that the only light degrees of freedom are precisely the ones needed for the MSSM with nothing more.  This is already a bit too restrictive because gauge mediated models \cite{Marsano:2008jq,Heckman:2008qt} suggest that we should incorporate vector-like pairs of messenger fields, $f$ and $\overline{f}$, which couple to an MSSM singlet, $X$, responsible for breaking supersymmetry via the coupling
\begin{equation}Xf\overline{f}\,.\end{equation}
In the $E_8$ scenarios of \cite{Bouchard:2009bu,Heckman:2009mn}, for instance, the fields $f$ and $\overline{f}$ transform in the $\mathbf{10}$ and $\mathbf{\overline{10}}$ of $SU(5)$.  As we describe in section \ref{sec:constraints}, however, to avoid the restricted class of spectral covers described in this paper we need something a bit more radical:  we need to allow a messenger sector in which $f$ and $\overline{f}$ which do not comprise complete GUT multiplets.

One might think that this is a problem for unification but, quite interestingly, F-theory GUTs have a complimentary puzzle as well.  As pointed out in \cite{Blumenhagen:2008aw}, the internal hypercharge flux used to break the $SU(5)$ gauge group already disrupts unification at the scale $M_{GUT}$.  This opens up a new puzzle, namely why experimental data indicates that the gauge couplings \emph{seem} to unify while, in this class of models, they actually fail to do so.  In section \ref{sec:constraints}, we will note that the incomplete GUT multiplets needed to avoid our constraints are of precisely the right form to effectively "cancel" the effects of this internal hypercharge flux.  To be sure, getting favorable neutrino physics forces us to introduce new degrees of freedom of just the right type to address the unification problem in F-theory GUTs.  We will have more to say about this in future work \cite{futurework}, where we hope to construct compact models of this type.

\subsection{Outline}

The outline of this paper is as follows.  In section \ref{sec:review}, we review the basic structure of elliptically fibered Calabi-Yau four-folds with resolved $E_8$ singularities as well as the symmetries that constrain the superpotential in such models.  We describe the origin of monodromies and review the spectral cover construction that is useful for studying them.  In section \ref{sec:constraints}, we discuss the specific constraints that we impose throughout our model-building efforts and demonstrate that a very limited number of scenarios can satisfy all of them.  We also comment on how to avoid these constraints by introducing a messenger sector comprised of incomplete GUT multiplets.  We study generic features of models that can realize most of our constraints in section \ref{sec:factorization} before proceeding to construct explicit 3-generation GUTs in section \ref{sec:compact}.
Finally, in section \ref{sec:mu} we comment on conditions that must be satisfied to address the $\mu$ problem in our compactifications.  Several computational details, as well as a brief review of the three-fold constructed in \cite{Marsano:2009ym} that serves as a base of our compact elliptically fibered Calabi-Yau four-folds, are deferred to the Appendices.



\section{Four-folds, Monodromies, and Fluxes}
\label{sec:review}

In this section, we review the generic structure of Calabi-Yau four-folds with resolved $E_8$ singularities and the effective 4-dimensional field theories that they engineer.  In the first two subsections, we recall basic aspects of the geometry and proceed to discuss the origin of various symmetries that constrain the 4-dimensional superpotential.  After that, we review the construction of well-defined $G$-fluxes necessary for engineering chiral matter.



\subsection{Structure of Four-folds}
\label{sec:Fourfold}

We begin by reviewing the structure of local Calabi-Yau four-folds for $SU(5)$ GUTs.  
We consider elliptically fibered Calabi-Yau four-folds, with base three-fold $B_3$.  For most of this paper, we focus our attention on four-folds which take the form of a local ALE fibration over a four-cycle, $S_{\rm GUT}$, over which the fiber degenerates to an $SU(5)$ singularity.  Charged matter fields and Yukawa couplings originate from curves and points where the singularity type enhances in rank.  Eventually we will also consider the embedding of such local four-folds into honest compact ones.

Our starting point is the Weierstrass model
\be
y^2 = x^3  + f x + g \,,
\ee
where
$f$ and $g$ are holomorphic sections of $(N_{S_{\rm GUT}/B_3}\otimes K_{S_{\rm GUT}}^{-1})^4$ and $(N_{S_{\rm GUT}/B_3}\otimes K_{S_{\rm GUT}}^{-1})^6$, respectively.  Here, $N_{S_{\rm GUT}/B_3}$ and $K_{S_{\rm GUT}}$ denote the normal bundle of $S_{\rm GUT}$ in $B_3$ and the canonical bundle of $S_{\rm GUT}$, respectively.

The surface $S_{\rm GUT}$ on which we seek to realize the $SU(5)$ gauge degrees of freedom is a holomorphic divisor inside $B_3$.  As such, it is specified by the vanishing of a holomorphic section, $z$, which provides for us a local coordinate for the "normal" direction to $S_{\rm GUT}$.  Expanding $f$ and $g$ in the section $z$ while requiring an $SU(5)$ singularity along $z=0$  as in \cite{Andreas:2009uf}, the Weierstrass model can be put into the form of a  generic deformation of an $E_8$ singularity to $SU(5)$ \cite{Andreas:2009uf,Donagi:2009ra,Marsano:2009ym}
\begin{equation}
\label{calabi}
y^2=x^3+b_5xy+b_4x^2z+b_3yz^2+b_2xz^3+b_0z^5\,,
\end{equation}
where $x$, $y$, and the $b_m$ are holomorphic sections of suitable bundles over $S_{\rm{GUT}}$.  We will use standard notation in which $c_1$ stands for the first Chern class of the tangent bundle to $S_{\rm{GUT}}$
and $-t$ for the first Chern class of the normal bundle of ${S_{\rm GUT}}$ inside the base $B_3$. With this notation, the objects appearing here are sections of line bundles as denoted in the following table
\begin{equation}\begin{array}{c|c}\text{Section} & c_1\left(\text{Bundle}\right) \\ \hline
y & 3(c_1-t) \\
x & 2(c_1-t) \\
z & -t\\
b_m & \eta - mc_1
\end{array}\label{bmsectinfo}\end{equation}
where
\begin{equation}\eta = 6c_1-t.\label{etadef}\end{equation}

\subsubsection{Origin of Charged Matter and Yukawa Couplings}

\begin{figure}
\begin{center}
\epsfig{file=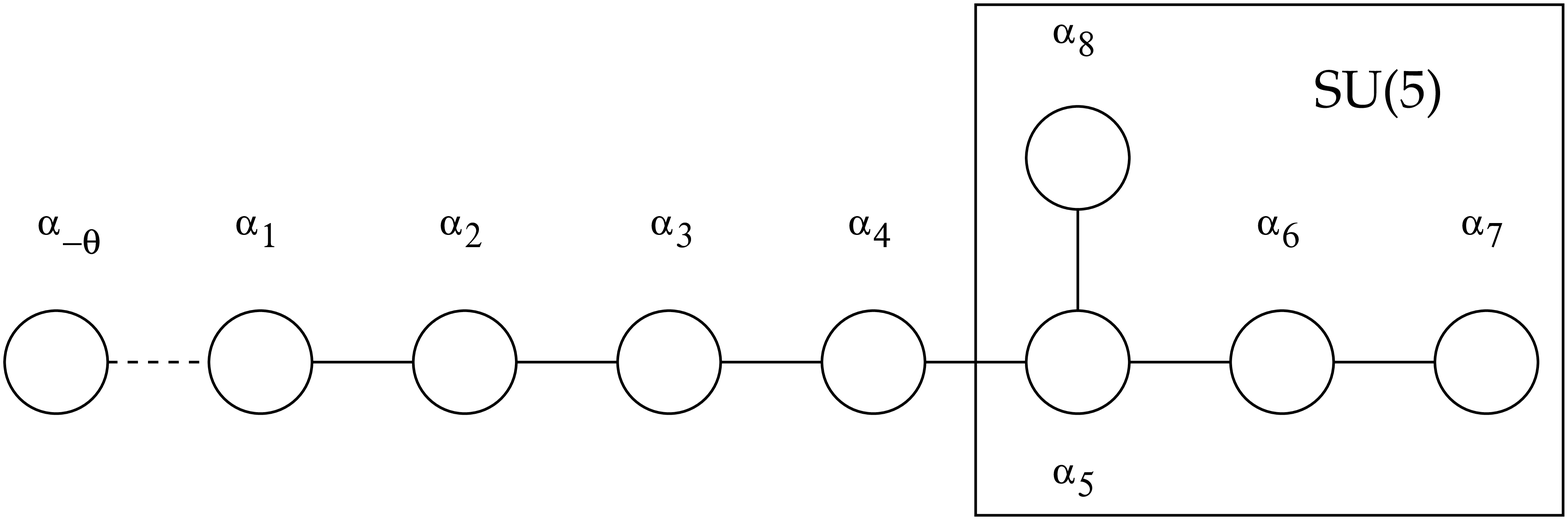,width=0.6\textwidth}
\caption{Extended $E_8$ Dynkin diagram}
\label{E8Dynkin}
\end{center}
\end{figure}

Let us take a moment to review the origin of charged matter and Yukawa couplings in geometries of this type.  A full $E_8$ singularity exhibits 8 collapsed $\mathbb{P}^1$'s whose intersection matrix is given by $-1$ times the $E_8$ Cartan matrix.  This provides a natural identification of geometric parameters, namely the volumes of the 8 $\mathbb{P}^1$'s, with the simple roots of $E_8$ associated to nodes of the corresponding Dynkin diagram.  In figure \ref{E8Dynkin}, we depict the extended $E_8$ Dynkin diagram which, in addition to the 8 nodes $\alpha_1,\ldots,\alpha_8$, includes also the node $\alpha_{-\theta}$
\begin{equation}\alpha_{-\theta}=-2\alpha_1-3\alpha_2-4\alpha_3-5\alpha_4-6\alpha_5-4\alpha_6-2\alpha_7-3\alpha_8\,.
\end{equation}

An $E_8$ singularity over $S_{\rm GUT}$ engineers an 8-dimensional $E_8$ gauge theory on $\mathbb{R}^{3,1}\times S_{\rm GUT}$.  Resolving the singularity by introducing nonzero volumes for some of the $\alpha_i$ corresponds to turning on a nontrivial expectation value for the adjoint scalar field of this theory, $\phi_{adj}$, that breaks $E_8$ to a subgroup.  We will be interested in $SU(5)$ models so we will take $\langle\phi_{adj}\rangle$ to lie in the Cartan subalgebra of the $SU(5)_{\perp}$ commutant of $SU(5)$ inside $E_8$.

At generic points, $\langle\phi_{\adj}\rangle$ will be such that it completely breaks $E_8\rightarrow SU(5)\times U(1)^4$.  Further, this expectation value gives a nonzero mass to several components of the $E_8$ adjoint chiral multiplet through the coupling
\begin{equation}[\langle \phi_{adj}\rangle,\delta\phi_{adj}]^2 \,.
\label{phiadjmass}\end{equation}
For typical $\langle\phi_{adj}\rangle$, only the $SU(5)$ adjoint and some $SU(5)$ singlets remain.
An easy way to motivate the relation between masses of the other residual $SU(5)$ multiplets and geometric volumes of $\mathbb{P}^1$'s is to recall the decomposition of the $E_8$ adjoint under $E_8\rightarrow SU(5)\times SU(5)_{\perp}$
\begin{equation}\mathbf{248}\rightarrow (\mathbf{24},\mathbf{1})\oplus (\mathbf{1},\mathbf{24})\oplus (\mathbf{10},\mathbf{5})\oplus(\mathbf{\overline{5}},\mathbf{10})\oplus (\mathbf{\overline{10}},\mathbf{\overline{5}})\oplus (\mathbf{5},\mathbf{\overline{10}}) \,.
\end{equation}
We see from this that multiplets transforming in the $\mathbf{10}$ of $SU(5)$ obtain masses from \eqref{phiadjmass} that correspond to the weights $\lambda_i$ of the fundamental $\mathbf{5}_{\perp}$ of $SU(5)_{\perp}$.
These weights are related to the roots $\alpha_i$, and hence the volumes of our $\mathbb{P}^1$'s, by the standard relations
\begin{equation}\begin{split}
\lambda_1 &= \alpha_4 \\
\lambda_2 &= \alpha_3+\alpha_4 \\
\lambda_3 &= \alpha_2+\alpha_3+\alpha_4 \\
\lambda_4 &= \alpha_1+\alpha_2+\alpha_3+\alpha_4 \\
\lambda_5 &= \alpha_{-\theta}+\alpha_1+\alpha_2+\alpha_3+\alpha_4 \,.
\end{split}\end{equation}
In general, then, the curves along which $\mathbf{10}$ and $\mathbf{5}$ matter fields localize are determined by
\begin{equation}\Sigma_{\mathbf{10}}\sim \lambda_i=0\,,\qquad\qquad \Sigma_{\mathbf{5}}\sim \lambda_j+\lambda_k=0,\,\,\, j\ne k\,.
\end{equation}
Note that we abuse notation a bit in letting the $\lambda_i$ refer both to weights of a given $SU(5)$ multiplet under $U(1)^4\subset SU(5)_{\perp}$ as well as the position-dependent masses whose vanishing specifies the matter curve on which that multiplet localizes.


Superpotential couplings involving these fields descend from the cubic $(\mathbf{248})^3$ term of the underlying $E_8$ gauge theory.  After the breaking $E_8\rightarrow SU(5)\times U(1)^4$, then, we expect to find all cubic couplings that are allowed by the $SU(5)\times U(1)^4$ symmetry.  The specific locus on which a field localizes, though uniquely determines its $U(1)^4$ charge.  For instance, a $\mathbf{10}$ associated to the curve $\lambda_1=0$ carries charge +1 with respect to the first $U(1)$ factor and 0 under the rest.  Similarly,  $\mathbf{\overline{5}}$ associated to the curve $\lambda_1+\lambda_2=0$ carries charge +1 with respect to the first two $U(1)$ factors and 0 under the rest.  A set of fields that can participate in up-type $\mathbf{10}\times\mathbf{10}\times\mathbf{5}$ Yukawas, then, must be associated to fields that localize on curves with $(\lambda_i=0)$, $(\lambda_j=0)$, and $(\lambda_i+\lambda_j=0)$ for some $i\ne j$.  The dominant contribution to such couplings will arise at the points $\lambda_i=\lambda_j=0$ where these curves meet.  For this reason, we associate up-type Yukawas with the points
\begin{equation}
\mathbf{10}_M\times\mathbf{10}_M\times\mathbf{5}_H\qquad\text{when}\qquad \lambda_i=\lambda_j=\lambda_i+\lambda_j=0\quad i\ne j \,.
\end{equation}
Similarly, we get down-type Yukawas from points where
\begin{equation}\mathbf{10}_M\times\mathbf{\overline{5}}_M\times\mathbf{\overline{5}}_H\qquad \text{when}\qquad \lambda_i+\lambda_j=\lambda_k+\lambda_{\ell}=\lambda_m=0\quad \epsilon_{ijk\ell m}\ne 0\,.
\end{equation}


\subsubsection{Matter Curves and Couplings in the Four-fold}

In the four-fold, \eqref{calabi}, one does not directly specify the volumes $\lambda_i$ but rather the objects $b_n$, to which they are related by \cite{Donagi:2009ra}
\begin{equation}b_n\sim b_0 e_n(\lambda_i)\,,
\end{equation}
where $e_n(\lambda_i)$ are the degree 5 elementary symmetric polynomials defined by
\begin{equation}\prod_{i=1}^5 (x+\lambda_i) = \sum_{n=0}^5 x^{5-n}e_n(\lambda_i)\,.\end{equation}
The absence of a $b_1$ term in \eqref{calabi} reflects the fact that $e_1(\lambda_i)\sim \sum_i\lambda_i=0$.  In terms of the $b_n$, the condition defining the $\mathbf{10}$ matter curve, namly $\lambda_i=0$ for some $i$, can be written as
\begin{equation}\Sigma_{\mathbf{10}}:\quad 0=b_5= \prod_{i=1}^5\lambda_i\,.\end{equation}
It is easy to see from \eqref{calabi} that this corresponds to an enhancement of the singularity type from $SU(5)$ to $SO(10)$.  On the other hand, the condition $\lambda_i+\lambda_j=0$ for some $i\ne j$ can be written as
\begin{equation}\Sigma_{\mathbf{5}}:\quad 0=\prod_{i<j}(\lambda_i+\lambda_j)=b_1b_2b_3b_4-b_1b_2^2b_5-b_1^2b_4^2+2b_0b_1b_4b_5+b_0(b_2 b_3b_5-b_3^2b_4-b_0b_5^2)\,.
\end{equation}
This means that, when $b_1=0$, $\mathbf{5}$ matter is localized along the curve given by
\begin{equation}P\equiv b_0 b_5^2 - b_2b_3b_5+b_3^2b_4=0\,.
\label{Pdef}\end{equation}
It is easy to verify from \eqref{calabi} that this corresponds to an enhancement of the singularity type from $SU(5)$ to $SU(6)$.

Turning now to superpotential couplings, the condition associated to up-type Yukawas, $\lambda_i=\lambda_j=\lambda_i+\lambda_j=0$ for some $i\ne j$, corresponds to
\begin{equation}\mathbf{10}\times\mathbf{10}\times\mathbf{5}:\quad b_3=b_5=0\end{equation}
This describes an $E_6$ enhancement of the singularity in \eqref{calabi}.  Similarly, the condition for down-type Yukawas, $\lambda_i+\lambda_j=\lambda_k+\lambda_{\ell}=\lambda_m=0$ for $i,j,k,\ell,m$ all distinct, is simply
\begin{equation}\mathbf{10}\times\mathbf{\overline{5}}\times\mathbf{\overline{5}}:\quad b_4=b_5=0\end{equation}
This describes an $SO(12)$ enhancement of the singularity in \eqref{calabi}.




\subsection{Monodromies and Symmetries}\label{subsec:monodsymm}

As described in the previous section, Yukawa couplings originate from the standard $\mathbf{248}\times\mathbf{248}\times\mathbf{248}$ cubic coupling of the underlying $E_8$ theory.  Because the breaking $E_8\rightarrow SU(5)$ reduces the rank of the gauge group by 4, one naively expects an additional 4 $U(1)$ gauge symmetries to remain and constrain the form of the superpotential.  Generically, however, this is not the case.  As we saw, the data of the four-fold does not specify distinct volumes, $\lambda_i$, but rather sections, $b_n$, to which the $\lambda_i$ are related by
\begin{equation}b_n\sim e_n(\lambda_i)\,.\end{equation}
In general, if we try to invert these equations to obtain $\lambda_i(b_n)$, the corresponding solutions will exhibit branch cuts.  The individual $\lambda_i(b_n)$ can only be defined locally and are subject to the action of a nontrivial monodromy group, $G$, that is a subgroup of the Weyl group $W_{A_4}\cong S_5$ of $SU(5)_{\perp}$.

The nature of the monodromy group $G$ has important implications both for the dynamical (but possibly anomalous) gauge symmetries that remain as well as the structure of the superpotential.  It played a crucial role in local studies of neutrino scenarios in \cite{Bouchard:2009bu,Heckman:2009mn} and also has implications for the types of fluxes that can be introduced for realizing three generation models \cite{Donagi:2009ra}.

\subsubsection{The Spectral Surface, ${\cal{C}}_{\mathbf{10}}$}

While all of the data of the monodromy group is contained in the $b_m$, it is often useful to introduce an auxiliary object, the fundamental spectral surface ${\cal{C}}_{\mathbf{10}}$, to describe it.  A convenient realization of this surface was given in \cite{Donagi:2009ra} as a submanifold of the projective three-fold
\begin{equation}X=\mathbb{P}({\cal{O}}_{S_{\rm GUT}}\oplus K_{S_{\rm GUT}})\,,\end{equation}
defined by the equation
\begin{equation}{\cal{F}}_{ {\cal{C}}_{\mathbf{10}}}=b_0 U^5 + b_2V^2U^3+b_3V^3U^2+b_4V^4U+b_5V^5=0\,.
\label{fund}\end{equation}
Here ${\cal{O}}_{S_{\rm GUT}}$ and $K_{S_{\rm GUT}}$ are the trivial and canonical bundle on $S_{\rm GUT}$, respectively.  The homogeneous coordiantes $[U,V]$ on the $\mathbb{P}^1$ fiber are sections of ${\cal{O}}(1)\otimes K_{S_{\rm GUT}}$ and ${\cal{O}}(1)$, respectively, where ${\cal{O}}(1)$ is the line bundle of degree 1 on $\mathbb{P}^1$.

It will also be convenient to define the projection
\begin{equation}\pi:X\rightarrow S_{\rm GUT}\end{equation}
along with the map, $p_{{\cC}_{10}}$, that it induces
\begin{equation}p_{{\cC}_{10}}:{\cal{C}}_{\mathbf{10}}\rightarrow S_{\rm GUT}\,.\label{pCdef}\end{equation}
The object ${\cal{F}}_{{\cal{C}}_{\mathbf{10}}}$ is a projectivization of the equation
\begin{equation}0=b_0s^5 + b_2s^3+b_3s^2+b_4s+b_5\sim b_0\prod_{i=1}^5 (s+\lambda_i)\,,\end{equation}
whose roots, as indicated, are essentially the $\lambda_i$.  In any local patch, the sheets of ${\cal{C}}_{\mathbf{10}}$ provide a solution $\lambda_i(b_n)$ while the monodromy group, $G$, is encoded by the topology of the full surface.


\subsubsection{Monodromies and the Factorization of ${\cal{C}}_{\mathbf{10}}$}
\label{monodclass}

The most immediate characteristic of ${\cal{C}}_{\mathbf{10}}$ is the number of components into which it factors.  This carries significant information about $G$ because it describes the orbits of the $\{\lambda_i\}$ under its action.  There are several possibilities, which we now describe.

\begin{itemize}

\item
\textbf{Case 1: ${\cal{C}}_{\mathbf{10}}$ does not factor}

If ${\cal{C}}_{\mathbf{10}}$ does not factor then $G$ is a transitive subgroup of $S_5$.  Up to conjugacy, there are only four proper transitive subgroups of $S_5$.  These include the alternating group of even permutations, $A_5$, the cyclic group of order 5, $\mathbb{Z}_5$, the dihedral group, $D_5$, and a group of order 20 generated by $(12345)$ and $(2354)$.

\item
\textbf{Case 2: ${\cal{C}}_{\mathbf{10}} = {\cal{C}}_{\mathbf{10}}^{(1)}+{\cal{C}}_{\mathbf{10}}^{(4)}$}

If ${\cal{C}}_{\mathbf{10}}$ splits into a linear and a quartic piece then the monodromy group is a transitive subgroup of $S_4$.  The only proper transitive subgroups of $S_4$ are
\begin{equation}A_4,D_4,\mathbb{Z}_4,V\end{equation}
where $A_4$ is the alternating group of even permutations, $D_4$ is a dihedral group, $\mathbb{Z}_4$ is the cyclic group of order 4, and $V\cong\mathbb{Z}_2\times\mathbb{Z}_2$ is the Klein four group generated by even permutations of order 2
\begin{equation}V=\{1,(12)(34),(13)(24),(14)(23)\}\,.
\end{equation}

\item
\textbf{Case 3: ${\cal{C}}_{\mathbf{10}}={\cal{C}}_{\mathbf{10}}^{(3)}+\ldots$}

If ${\cal{C}}_{\mathbf{10}}$ contains a cubic piece then the monodromy group is a transitive subgroup of either $S_3\times S_2$ or $S_3$, depending on whether or not the remaining degree 2 piece factors.  Note that $S_2$ has no nontrivial proper subgroups while the only proper transitive subgroup of $S_3$ is $A_3\cong \mathbb{Z}_3$.  The monodromy group in this case is either $S_3\times S_2$, $A_3\times S_2$, $S_3$, or $A_3$.

\item
\textbf{Case 4: ${\cal{C}}_{\mathbf{10}}=\sum_i {\cal{C}}_{\mathbf{10}}^{(m_i)}$ for $m_i\le 2$}

If ${\cal{C}}_{\mathbf{10}}$ splits into only linear and quadratic factors then $G$ is completely determined as a product of the relevant $\mathbb{Z}_2$'s.

\end{itemize}

As we shall see below, the specific monodromy group $G$ in a given case is often crucial in determining the structure of the superpotential.

\subsubsection{Matter Curves in ${\cal{C}}_{\mathbf{10}}$}
\label{subsubsec:mattcurvC10}

An important observation of \cite{Tatar:2009jk} is that when two matter curves meet without a further enhancement in the singularity type of the fiber, their wave functions are not independent but instead are connected by a nontrivial boundary condition at the intersection point.  This happens because the $SU(5)$ multiplets on the two curves locally exhibit an identical embedding into $E_8$, meaning that they arise from a single wave function on $S_{\rm GUT}$ that happens to localize on both pieces.  When counting zero modes, it is more appropriate to think of such naively independent matter curves as a single one that happens to "pinch" at one or more points.  Even though the different pieces that remain after this "pinching" are topologically split, the wave functions see them as a single object so we will always treat them as such.  In particular, we will always use the term \emph{matter curve} to refer only to the entire (possibly reducible) curve on which a given $(SU(5)\times U(1)^4)/G$ multiplet is localized{\footnote{In some special cases, the matter curve associated with a single multiplet might split into pieces in such way that one piece, $\tilde{\Sigma}$, fails to intersect all of the others.  In that case, one should be able to distinguish zero modes on $\tilde{\Sigma}$ and zero modes on the rest.  This situation is somewhat nongeneric so we will not consider it in this paper.}}

For trivial monodromy group, we would expect at most 5 independent $\mathbf{10}$ matter curves, corresponding to the five copies of the $\mathbf{10}$ inside the $E_8$ adjoint and defined by the five equations $\lambda_i=0$.  When various $\lambda_i$ are mixed by monodromies, however, the number of independent matter curves decreases. A useful way to keep track of this is by visualizing the $\mathbf{10}$ matter curves directly inside the spectral surface $\mathcal{C}_{\mathbf{10}}$.  This is done by noting that a $\mathbf{10}$ matter curve corresponds to the intersection of any sheet $s\sim \lambda_i$ with zero so that, inside $X$, it is given by the intersection
\begin{equation}\Sigma_{\mathbf{10}}\sim (U=0)\cap \mathcal{C}_{\mathbf{10}}\end{equation}
Two curves whose $SU(5)$ multiplets are related by monodromies will lie in a common component of $\mathcal{C}_{\mathbf{10}}$.  On the other hand, two curves whose $SU(5)$ multiplets are not related by monodromies will lie in different components of $\mathcal{C}_{\mathbf{10}}$.

For $\mathbf{\overline{5}}$ matter fields, the situation is slightly more tricky.  Often one introduces another surface, ${\cal{C}}_{\overline{\mathbf{5}}}$, by projectivizing \cite{Donagi:2009ra}
\begin{equation}\begin{split}\prod_{i<j}(s+\lambda_i+\lambda_j)&\sim b_0^3\left[s^{10}+3s^8c_2-s^7c_3+3s^6(c_2^2-c_4)+s^5(-2c_2c_3+11c_5)+s^4(c_2^3-c_3^2-2c_2c_4)\right.\\
&\qquad + s^3(-c_2^2c_3+4c_3c_4+4c_2c_5)+s^2(-c_2c_3^2+c_2^2c_4-4c_4^2+7c_3c_5)\\
&\qquad\left.+s(c_3^3c_2^2c_5-4c_4c_5)+(c_2c_3c_5-c_3^2c_4-c_5^2)\right] \,.
\end{split}\end{equation}
where $c_m=b_m/b_0$.  In that case, the net $\mathbf{\overline{5}}$ matter curve is simply described as the intersection of $s=0$ with ${\cal{C}}_{\overline{\mathbf{5}}}$.  Different pieces of this curve correspond to truly distinct matter curves precisely when they lie in different components of ${\cal{C}}_{\overline{\mathbf{5}}}$.  It is often useful, however, to visualize the $\mathbf{\overline{5}}$ matter curves directly in ${\cal{C}}_{\mathbf{10}}$.  This allows us to associate them with specific sheets of $\mathcal{C}_{{\bf 10}}$, helping us to identify the local "charges" of both $\mathbf{10}$'s and $\mathbf{\overline{5}}$'s under the $U(1)^4$ Cartan of $SU(5)_{\perp}$ at the same time.

To obtain a prescription for picking out the $\mathbf{\overline{5}}$ matter curve inside ${\cal{C}}_{\mathbf{10}}$, we follow the heterotic literature (see for instance \cite{Donagi:2004ia,Blumenhagen:2006wj,Hayashi:2008ba}) and define an involution $\tau:X\rightarrow X$ by sending $V\rightarrow -V$, which inverts the sheets of ${\cal{C}}_{\mathbf{10}}$ according to $\lambda_i\rightarrow -\lambda_i$.  The fixed locus of ${\cal{C}}_{\mathbf{10}}$ under this involution consists of three components
\begin{itemize}
\item $\lambda_i=0$
\item $\lambda_i+\lambda_j=0$
\item $\lambda_i\rightarrow \infty$
\end{itemize}
The first component is the $\mathbf{10}$ matter curve while the third is the intersection (of appropriate multiplicity) of ${\cal{C}}_{\mathbf{10}}$ with the "divisor at infinity" along $V=0$.  In the heterotic literature, where spectral surfaces play a crucial role, this component is variously referred to as ${\cal{C}}_V\cap \sigma_2$ in \cite{Donagi:2004ia} or $C\cap \sigma_t$  in \cite{Blumenhagen:2006wj}.  The third component is precisely our desired $\mathbf{\overline{5}}$ matter curve.

\subsubsection{Imprint of $G$ on the Superpotential}\label{subsubsec:imprint}

The most obvious impact of $G$ on the form of the superpotential is in the collection of $U(1)$ gauge symmetries that remain after we perform the quotient.  Typically,
of the four $U(1)$ gauge symmetries that arise when $E_8$ is broken to $SU(5)$, all of them are projected out by the monodromy group action.  To see this, recall that the gauge boson associated to such a symmetry corresponds to an element of the Cartan subalgebra, and thus to the dual space to the $\lambda_i$, which can be written as a linear 
combination
\begin{equation}\sum_i c_i \lambda_i^{\ast}\,,
\end{equation}
where $\lambda_i^{\ast}(\lambda_j)=\delta_{ij}$.  The only linear combination of $\lambda_i$ that is invariant under the generic monodromy group, $S_5$, is $\sum_i\lambda_i$, namely the one linear combination that vanishes identically.  In fact, the same is true for any subgroup of $S_5$ that acts transitively on the $\lambda_i$.  In order to preserve a dynamical $U(1)$ gauge symmetry, then, it is necessary for the set of sheets, $\{\lambda_i\}$, to comprise a reducible representation of $G$.  In terms of the spectral surface, this is equivalent to the statement that ${\cal{C}}_{\mathbf{10}}$ factors into distinct components.  The mondromy group $G$ in such a situation is contained within the suitable product of symmetric groups and the number of surviving $U(1)$ gauge symmetries is $N-1$, where $N$ is the number of components of ${\cal{C}}_{\mathbf{10}}$.

Sometimes, however, these $U(1)$ symmetries alone are not sufficient to understand all of the structure that is inherited from the underlying $E_8$ "parent" theory.  Consider, for instance, a situation in which $G=\mathbb{Z}_4$ generated by the element $(1234)$ acting on the $\lambda_i$ with $i=1,\ldots,4$.  In this case, there are two $\mathbf{10}$ and three $\mathbf{\overline{5}}$ matter curves, obtained by decomposing the $\lambda_i$ and $\lambda_i+\lambda_j$ into orbits of $G$, as
\begin{equation}\begin{split}\mathbf{10}^{(1)} &= \{\lambda_1,\lambda_2,\lambda_3,\lambda_4\} \\
\mathbf{10}^{(2)} &= \{\lambda_5\} \\
\mathbf{\overline{5}}^{(1)} &\sim \{\lambda_1+\lambda_2,\lambda_2+\lambda_3,\lambda_3+\lambda_4,\lambda_4+\lambda_1\} \\
\mathbf{\overline{5}}^{(2)} &= \{\lambda_1+\lambda_3,\lambda_2+\lambda_4\} \\
\mathbf{\overline{5}}^{(3)} &= \{\lambda_1+\lambda_5,\lambda_2+\lambda_5,\lambda_3+\lambda_5,\lambda_4+\lambda_5\}\,.
\end{split}\end{equation}
All superpotential couplings must not only be $SU(5)$ invariant but must also descend from couplings that are invariant under the $U(1)^4$ left over in the decomposition $E_8\rightarrow SU(5)\times U(1)^4$ before we quotient by $G$.  More specifically, $\mathbf{10}\times\mathbf{10}\times\mathbf{5}$ and $\mathbf{10}\times\mathbf{\overline{5}}\times\mathbf{\overline{5}}$ couplings arise from points of $E_6$ and $SO(12)$ enhancement, respectively, which are described by
\begin{equation}\begin{split}
E_6:& \qquad  \lambda_i=\lambda_j=\lambda_i+\lambda_j=0\,,\qquad i\ne j \\
SO(12):&\qquad  \lambda_i=\lambda_j+\lambda_k=\lambda_{\ell}+\lambda_m=0\,,\qquad \epsilon_{ijklm}\ne 0\,.
\end{split}\end{equation}
This means that the several couplings are allowed, including
\begin{equation}\mathbf{10}^{(1)}\times\mathbf{10}^{(1)}\times\mathbf{5}^{(1)},\quad \mathbf{10}^{(1)}\times\mathbf{10}^{(1)}\times\mathbf{5}^{(2)},\quad\mathbf{10}^{(2)}\times\mathbf{\overline{5}}^{(1)}\times\mathbf{\overline{5}}^{(1)},\quad\mathbf{10}^{(2)}\times\mathbf{\overline{5}}^{(2)}\times\mathbf{\overline{5}}^{(2)}\label{allowed}\end{equation}
but the following is forbidden
\begin{equation}\mathbf{10}^{(2)}\times\mathbf{\overline{5}}^{(1)}\times\mathbf{\overline{5}}^{(2)} \,.
\label{forbidden}\end{equation}
There is no symmetry in the "daughter" theory after the quotient that can forbid \eqref{forbidden} while allowing all of the couplings in \eqref{allowed}.  The fact that \eqref{forbidden} is forbidden is a reflection of the monodromy group and an indication that simply knowing the number of components of ${\cal{C}}_{\mathbf{10}}$ is not enough to determine the structure of the theory.

So, when do we have to worry about $U(1)$'s not being sufficient?  In general, to classify all possible Yukawa couplings involving $\mathbf{10}$'s and $\mathbf{\overline{5}}$'s (and their conjugates), it is necessary to determine the decomposition of $\{\lambda_i\}$ and $\{\lambda_i+\lambda_j\}$ into orbits under the action of $G$.  The former can be obtained from knowledge of the components of ${\cal{C}}_{\mathbf{10}}$.  Likewise, the latter can be determined from the number of components into which ${\cal{C}}_{\mathbf{\overline{5}}}$ splits.  From the classification of section \ref{monodclass}, however, the monodromy group is completely specified when ${\cal{C}}_{\mathbf{10}}$ contains only linear and quadratic factors.  Further, when ${\cal{C}}_{\mathbf{10}}$ contains a cubic factor the only ambiguity is whether the group associated to that factor is $S_3$ or $A_3$.  Note, however, that $A_3$ acts transitively on the set $\{\lambda_a+\lambda_b\}$ for $a,b=1,2,3$.  For purposes of determining superpotential couplings involving $\mathbf{10}$'s and $\mathbf{\overline{5}}$'s, then, the $S_3$ and $A_3$ cases are indistinguishible{\footnote{If we are interested in couplings involving singlet fields, though, the two cases would lead to different structures.}}.  As such, we see that \emph{if ${\cal{C}}_{\mathbf{10}}$ contains no components of degree 4 or 5 then all couplings are allowed except those that are expressly  forbidden by residual $U(1)$ gauge symmetries}.

\subsection{$G$-fluxes from spectral cover}
In addition to specifying a four-fold, we must describe the various $G$-fluxes that are turned on.  Because we do not address moduli stabilization in this paper, we will focus attention only on those $G$-fluxes that are relevant for determining the spectrum of charged matter.  To see which ones we need recall that, from the M-theory viewpoint, the charged fields on matter curves correspond to M2 branes wrapping the 2-cycles that degenerate there.  The $G$-fluxes to which these M2 branes couple are of the form
\begin{equation}G = \omega_i\wedge F_i \,,
\label{Gfluxform}\end{equation}
where $\omega_i$ are harmonic (1,1) forms satisfying
\begin{equation}\int_{\lambda_j}\omega_i=\delta_{ij}\,.\end{equation}
A $G$-flux of the form \eqref{Gfluxform}, therefore, can be interpreted as a flux $F_i$ in the $U(1)$ subgroup of $E_8$ corresponding to $\omega_i$.

To describe the $G$-fluxes that we can turn on, then, it is important to choose a basis for the $\omega_i$ or, equivalently, a basis for the 2-cycles that degenerate on various matter curves.  A natural choice is provided by the 5 $\lambda_i$'s, which satisfy one nontrivial relation
\begin{equation}\sum_i \lambda_i=0\,.\end{equation}

Of course, it is necessary to properly account for the monodromy structure captured by ${\cal{C}}_{10}$.  According to the discussion of \cite{Donagi:2009ra}, the proper way to do this is derive $F$ from a well-defined line bundle, $L$, on ${\cal{C}}_{10}$ that satisfies
\begin{equation}0 = c_1(p_{ {\cal{C}}_{10}\ast}L) = p_{ {\cal{C}}_{10}\ast}c_1(L) - \frac{1}{2}p_{ {\cal{C}}_{10}\ast }r\,,\end{equation}
where $r$ is the ramification divisor
\begin{equation}r=p_{ {\cal{C}}_{10}}^\ast c_1(S_{\rm GUT}) - c_1({\cal{C}}_{10})\,.\end{equation}
It is natural to decompose $c_1(L)$ as
\begin{equation}c_1(L) = \frac{1}{2}r + \gamma\,,\end{equation}
where $\gamma$ satisfies
\begin{equation}p_{ {\cal{C}}_{10}\ast}\gamma=0\,.\end{equation}
Our desired flux $F$ is nothing other than this object $\gamma$, which must be quantized in such a way that it is consistent with $L$ being an integer line bundle.


For generic $b_0,\ldots,b_5$, there is only one independent class $\gamma_u$, the so-called universal class.
Below we review how to identify matter curves and determine the chiral spectrum that arises from $\gamma_u$ \cite{Donagi:2009ra}.  In section 4 we will discuss the chiral spectrum from certain non-universal fluxes in a non-generic
situation in which the spectral surface decomposes into linear and quartic surfaces, ${\cal{C}}_{10}={\cal{C}}_{10}^{(1)}+{\cal{C}}_{10}^{(4)}$.

\subsubsection{Matter Curves}

To determine the chiral spectrum induced by a given $G$-flux, it is necessary to integrate $\gamma$ over matter curves in ${\cal{C}}_{10}$.  This requires a more specific analysis of the structure of matter curves in ${\cal{C}}_{10}$, which we review in this section.

First, however, let us make a few remarks about divisors in the ambient projective bundle $X$ into which ${\cal{C}}_{10}$ is embedded.  Recall that
\begin{equation}X=\mathbb{P}({\cal{O}}_{S_{\rm GUT}}\oplus K_{S_{\rm GUT}})\,.\end{equation}
The divisor $U=0$ corresponds to the zero section, $\sigma$, which is the class of $S_{\rm GUT}$ in $H_4(X,\mathbb{Z})$.  On the other hand, $V=0$ corresponds to the "divisor at infinity", which we shall denote by $\sigma_{\infty}$.  Using \eqref{bmsectinfo} and \eqref{etadef}, this implies that
\begin{equation}\sigma_{\infty}=\sigma+c_1\,,\end{equation}
and that the net class of ${\cal{C}}_{10}$ inside $X$ is simply
\begin{equation}[{\cal{C}}_{\bf 10}] = 5\sigma + \pi^{\ast}\eta\,.\end{equation}
Further, because the intersection of $U=0$ with $V=0$ is empty, we have that $\sigma\cdot\sigma_{\infty}=0$ or, equivalently, that
\begin{equation}\sigma^2 = -\sigma\cdot c_1\,.\end{equation}


Let us now turn to the matter curves.  Recall from the discussion of section \ref{subsubsec:mattcurvC10} that both the $\mathbf{10}$ and $\mathbf{\overline{5}}$ matter curves are contained in the fixed locus of ${\cal{C}}_{10}$ under $\lambda_i\rightarrow -\lambda_i$ and hence under the involution
\begin{equation}\tau:V\rightarrow -V\,.
\end{equation}
Using the fact that $[\tau {\cC}_{10}] = [{\cC}_{10}]$, we can determine the homological class of this intersection inside $X$ from
\begin{equation}{{\cC}}_{10}\cdot {{\cC}}_{10} = 25\sigma^2 + 10 \sigma\cdot\pi^{\ast}\eta + (\pi^{\ast}\eta)^2\,.\end{equation}
At the moment, however, this is not very useful for us.  Let us instead try to compute this intersection directly from \eqref{fund}.  This intersection corresponds to simultaneous solutions to the two equations
\begin{equation}\begin{split}0 &= V^3\left(b_3U^2 + b_5 V^2\right) \\
0 &= U \left(b_0U^4 + b_2U^2V^2+b_4V^4\right)\,.
\label{twoeqns}\end{split}\end{equation}

There are two obvious components of interest.  The first is $U=0$, which appears as a single root of the second equation of \eqref{twoeqns}.  The intersection of $U=0$ with $F_{\mathcal{C}_{10}}$ is in the class
\begin{equation}{\cal{C}}_{10}\cap \sigma\,.
\end{equation}
This is the class of the $\mathbf{10}$ matter curve, $\Sigma_{10}$, inside ${\cal{C}}_{10}$ as it corresponds to the locus $U=0=b_5$.

The second component of interest is $V=0$, which appears as a triple root of the first equation of \eqref{twoeqns}.  This is the intersection of ${\cC}_{10}$ with the "divisor at infinity" and occurs with a three-fold degeneracy.  The class of this intersection is
\begin{equation}{\cC}_{10}\cap 3\sigma_{\infty} = {\cC}_{10} \cap 3\left(\sigma+\pi^{\ast}c_1\right)\,.\end{equation}
This corresponds to the $\lambda_i\rightarrow\infty$ locus of section \ref{subsubsec:mattcurvC10} that we are instructed to discard.

What remains, now, is the intersection
\begin{equation}\left({\cC}_{10}-U\right)\cap \left({\cC}_{10}-3V\right) = {\cC}_{10}\cap \left({\cC}_{10}-U-3V\right)\,.\end{equation}
The LHS here represents the precise intersection as what we have left are simultaneous solutions to
\begin{equation}\begin{split}0&= b_3U^2+b_5V^2 \\
0 &= b_0U^4 + b_2U^2V^2+b_4V^4\,.\end{split}\label{twoeqnsprime}\end{equation}
However, we have used the fact that $U\cap V=0$ to rewrite it as the intersection of a divisor in $X$ with ${\cC}_{10}$.  In particular, we note that
\begin{equation}{\cC}_{10}-U-3V = \sigma + \pi^{\ast}(\eta - 3c_1)\,.
\end{equation}
This will be useful later.  For now, let us note that this component is the class of the matter curve inside ${\cC}_{10}$ on which the $\mathbf{5}$ matter fields are localized.  To see that it projects to the curve $P=0$ \eqref{Pdef} inside $S_{\rm GUT}$, let us try to solve \eqref{twoeqnsprime}.  If we suppose that $b_3\ne 0$ then the first equation gives us
\begin{equation}U=\pm i\sqrt{\frac{b_5}{b_3}}V\,.\label{UVsol}\end{equation}
Plugging into the second equation yields
\begin{equation}\frac{V^4}{b_3^2}P=0\,,\qquad\qquad P\equiv b_0 b_5^2-b_2b_3b_5+b_3^2b_4\,.\end{equation}
This is a two-sheeted cover of the $\mathbf{5}$ matter curve inside ${S_{\rm GUT}}$.
We must also include the "points at infinity" where $b_3=V=0$.  Because these points are in ${\cC}_{10}$, they necessarily have $b_0=0$ so that they lie above the locus $b_3=b_0=0$ in ${S_{\rm GUT}}$.

Homologically, we compute
\begin{equation}{\cC}_{10}\cap \left({\cC}_{10}-U-3V\right) = 2\times \sigma\cdot \pi^{\ast}\left(3\eta-10c_1\right) + \pi^{\ast}\eta\cdot \pi^{\ast}\left(\eta-3c_1\right)\,.\end{equation}
We recognize $3\eta-10c_1$ as the class of $P=0$ inside ${S_{\rm GUT}}$.  The remaining term is a homological "correction" that accounts for the "points at infinity".

We can summarize these results in the following table

\begin{equation}\begin{array}{c|c}\text{Component} & \text{Class} \\ \hline
\mathbf{10} \text{ Matter Curve} & {\cal{C}}_{10}\cap \sigma \\
\mathbf{5}\text{ Matter Curve} & {\cal{C}}_{10} \cap \left(\sigma+\pi^{\ast}\eta-3\pi^{\ast}c_1\right) \\
\text{Intersection at infinity} & {\cal{C}}_{10}\cap 3\sigma_{\infty}
\end{array}\label{Cmattcurvs}\end{equation}


\subsubsection{Chiral Spectrum from universal flux}

The available fluxes are the traceless ones.  That is, they correspond to curves $\gamma$ inside ${\cal{C}}_{10}$ which satisfy
\begin{equation}p_{{\cC}_{10}\,\ast}\gamma=0\,.\label{traceless}\end{equation}
The "universal" $\gamma$ is given by
\begin{equation}\gamma_u = 5[\Sigma_{10}]_{{\cal{C}}_{10}}-p_{{\cC}_{10}}^{\ast}p_{{\cC}_{10}\,\ast}[\Sigma_{10}]_{{\cal{C}}_{10}}\,,\end{equation}
where we have indicated that $\Sigma_{10}$ is to be thought of as a class in ${{\cC}_{10}}$.  More precisely, from \eqref{Cmattcurvs} we see that $[\Sigma_{10}]_{{\cC}_{10}}$ is simply
\begin{equation}[\Sigma_{10}]_{{\cC}_{10}} = {{\cC}_{10}}\cap \sigma\,,
\end{equation}
where the intersection is in $X$.  The pushforward, $p_{C\,\ast}[\Sigma_{10}]_{{\cC}_{10}}$, is simply the class of $\Sigma_{10}$ inside ${S_{\rm GUT}}$, namely $\eta-5c_1$.  This means that
\begin{equation}\gamma_u = {{\cC}_{10}}\cap \left(5\sigma - \pi^{\ast}\eta +5\pi^{\ast}c_1\right)={{\cC}_{10}}\cap \left(5\sigma_{\infty}-\pi^{\ast}\eta\right)\,.\end{equation}
Now, the net chirality on the $\mathbf{10}$ matter curve is obtained by intersecting $\gamma_u$ with $[\Sigma_{10}]_{{\cC}_{10}}={{\cC}_{10}}\cap \sigma$ inside ${{\cC}_{10}}$.  This is simply
\begin{equation}\begin{split}n_{\mathbf{10}}-n_{\mathbf{\overline{10}}} &= {{\cC}_{10}}\cap \left(5\sigma-\pi^{\ast}\eta+5\pi^{\ast}c_1\right)\cap \sigma \\
&= {{\cC}_{10}}\cap \left(5\sigma_{\infty}-\pi^{\ast}\eta\right)\cap \sigma \\
&= -\pi^{\ast}\eta \cap {{\cC}_{10}}\cap \sigma \\
&= -\eta\cdot_{S_{\rm GUT}} \left(\eta-5c_1\right)\,.\end{split}\end{equation}

The net chirality of $\mathbf{\overline{5}}$'s is also easy to obtain.  In particular, we want to compute the intersection of $\gamma$ with the restriction of $\sigma+\pi^{\ast}\eta - 3\pi^{\ast}c_1$ to ${{\cC}_{10}}$.  Written as an intersection in ${{\cC}_{10}}$, this is
\begin{equation}\gamma\cdot_{{\cC}_{10}} \left(\sigma|_{{\cC}_{10}} + p_C^{\ast}\eta  - 3p_C^{\ast}c_1\right)\,.\end{equation}
We wrote everything this way to explicitly demonstrate the well-known fact that \emph{the net chirality of $\overline{\mathbf{5}}$'s always agrees with that of the $\mathbf{10}$'s}.  This follows because any $\gamma$ satisfying the traceless condition \eqref{traceless} also satisfies

\begin{equation}\gamma \cdot_{{\cC}_{10}} p_C^{\ast}\alpha=0\,,\qquad\forall\alpha\in H^2({S_{\rm GUT}},\mathbb{Z})\,.\end{equation}
This means that the net chirality of $\mathbf{\overline{5}}$'s is simply the intersection of $\gamma$ with $\sigma|_{{\cC}_{10}}$.  This is the same result that we obtained for the $\mathbf{10}$'s.  Note that this argument is sufficiently general that it will clearly hold in less generic situations when ${\cal{C}}_{10}$ factors.  We will make significant use of this fact later.

It is also instructive to perform this computation for the universal $\gamma$ by directly evaluating
\begin{equation}\begin{split}{{\cC}_{10}}\cap ({{\cC}_{10}}-U-3V) &\cap (5\sigma-\pi^{\ast}\eta+5\pi^{\ast}c_1) \\
&=\left[\sigma\cdot \pi^{\ast}(3\eta-10c_1)+\eta\cdot_{S_{\rm GUT}}(\eta-3c_1) F\right]\cdot \left(5\sigma-\pi^{\ast}\eta+5\pi^{\ast}c_1\right)\\
&= -\sigma\cdot \pi^{\ast}\eta\cdot\pi^{\ast}(3\eta-10c_1) + 2\eta\cdot_{S_{\rm GUT}} (\eta-3c_1) \\
&= -\eta\cdot_{S_{\rm GUT}} \left(\eta-5c_1\right)\,.
\end{split}\end{equation}

The above considerations were homological. In Appendix \ref{universal} we reproduce this result by
doing the computation with a specific representative of the universal $\gamma,$ which is useful
when $P$ factorizes.

\section{Constraints and $U(1)$ Symmetries}
\label{sec:constraints}

One of our primary goals is the construction of compact F-theory GUT models with as many realistic features as possible.  Before proceeding to discuss compact models, however, it is important to first spell out some simple requirements that we shall impose in order to lay the groundwork for obtaining realistic phenomenology.  These constraints are largely taken from the extensive literature on local models, though they comprise only a subset of the structures that have been introduced in that context.  Ultimately, we will find that a seemingly mild set of constraints severely restricts the class of constructions that are allowed.  The resulting models are guaranteed to have a gauged $U(1)$ that is a linear combination of $U(1)_{B-L}$ and $U(1)_Y$ and an accidental global $U(1)_{PQ}$ symmetry that emerges at the level of the renormalizable Lagrangian.

Because the resulting class of models is so restrictive, it is also interesting to consider the possibility of relaxing some of our constraints.  As we shall see, a constraint associated with the removal of certain charged exotics is of particular importance and we will comment on one possible way to evade it.  This scenario does not come without its own problems, though, so we will not pursue it further in the rest of this paper.


\subsection{Constraints}

Let us begin with a discussion of the constraints that we seek to impose when constructing F-theory GUT models.  Roughly, we can summarize them as follows.  We aim to achieve models with
\begin{itemize}
\item GUT-breaking and doubet-triplet splitting via nontrivial hypercharge flux
\item MSSM Superpotential
\item Absence of dangerous dimension 4 proton decay operators
\item Symmetries that forbid a bare $\mu$ term
\item Structure that favors the existence of flavor hierarchies
\end{itemize}
The issues of GUT-breaking and doublet-triplet splitting have been discussed at length in the literature \cite{Beasley:2008kw,Donagi:2008kj}.  What we need for this is an internal flux, $F_Y$, that threads the Higgs matter curves and is dual in $S_{\rm GUT}$ to two-cycle that is trivial in $B_3$.

\subsubsection{Dimension 4 Proton Decay Operators}

Before the appearance of \cite{Tatar:2009jk}, it was conventionally assumed that when a curve of enhanced symmetry factored inside $S_{\rm GUT}$, the wave functions that localized on the distinct factors were completely distinct.  This meant that the superpotential was completely determined by the intersection structure of the various factors so that, if we decided to localize $\mathbf{\overline{5}}_M$ on one factor and $\mathbf{10}_M$ on a second, the failure of these two curves to exhibit the proper intersection could be used to prevent the generation of a $\mathbf{10}_M\times\mathbf{\overline{5}}_M\times\mathbf{\overline{5}}_M$ superpotential coupling.

As reviewed in section \ref{subsubsec:mattcurvC10}, though, a crucial observation of \cite{Tatar:2009jk} is the inability to completely separate wave functions on factors of $\mathbf{10}$ ($\mathbf{\overline{5}}$) matter curves that come from the same component of ${\cal{C}}_{10}$ (${\cal{C}}_{5}$).  Such factors will typically intersect one another at a collection of points that do not exhibit a singularity of rank larger than 6.  In  that case, they comprise a single matter curve that has effectively "pinched" so that their wave functions are related through nontrivial boundary conditions.  It is for this reason that we reserve the term \emph{matter curve} for the complete locus on a component of ${\cal{C}}_{10}$ (${\cal{C}}_{5}$) where $\mathbf{10}$ ($\mathbf{\overline{5}}$) fields can localize.

What this means, however, is that the symmetry structure described in section \ref{subsec:monodsymm} is the only control that we have over the superpotential{\footnote{It was suggested in \cite{Tatar:2009jk} that some undesirable superpotential couplings could be suppressed if, say, the zero mode wave functions associated to $\mathbf{\overline{5}}_M$ and $\mathbf{\overline{5}}_H$ were realized on a single matter curve that "pinches" into factors but are effectively "localized" on different factors.  If this localization property can be realized, then the $\mathbf{10}_M\times\mathbf{\overline{5}}_M\times\mathbf{\overline{5}}_M$, while nonzero, would be significantly suppressed.  The authors of \cite{Tatar:2009jk} concluded that this suppression would likely not be sufficient to avoid conflicts with current bounds on the proton lifetime.}}.
To avoid running into trouble with current bounds on the proton lifetime, then, we must realize enough structure to expressly forbid the usual problematic dimension 4 operators, $\mathbf{10}_M\times\mathbf{\overline{5}}_M\times\mathbf{\overline{5}}_M$ and  $\mathbf{10}_M\times\mathbf{\overline{5}}_H\times\mathbf{\overline{5}}_H$.

\subsubsection{Favorable Flavor Structure}




The requirement that models contain a "structure that favors the existence of flavor hierarchies" is rather vague and requires further precision.  What we aim to achieve are hierarchies in the Yukawa matrices that arise in the manner proposed in \cite{Heckman:2008qa}, namely through the natural properties of wave function overlap integrals.  A necessary condition for this mechanism to operate is that all three generations of the $\mathbf{10}$ ($\mathbf{\overline{5}}$) must be engineered on a single $\mathbf{10}$ ($\mathbf{5}$) matter curve, $\Sigma_{10_M}$ ($\Sigma_{\overline{5}_M}$).

It is important to note that this condition is far from sufficient.  Indeed, to get honest hierarchies with minimal mixing in the quark sector, it is necessary that both the up-type and down-type Yukawa couplings originate from a \emph{single} unique point where the singularity type is enhanced to at least $E_7$. For the neutrino models of \cite{Heckman:2009mn}, it is in fact required to enhance up to $E_8$.
 At present, we are not so ambitious as to include this condition in our list of requirements.  Rather, we view this as a tuning that will be necessary to impose once certain weaker conditions are met.

\subsubsection{No Charged Exotics}

Finally, we require that there are no charged exotics.  In general, charged exotics that do not comprise a full GUT multiplet can arise whenever hypercharge flux threads a matter curve (other than the $\pm 1$ unit of hypercharge flux required to engineer Higgs doublets).
This is potentially troublesome in the case of $\mathbf{10}$ matter curves because we can associate to each a unique set of charges under the $U(1)$ gauge symmetries that remain after performing the quotient.  Exotic chiral matter fields on different $\mathbf{10}$ matter curves will therefore be unable to form invariant mass couplings, making it impossible to lift them from the theory{\footnote{It is possible that masses for these exotics get generated when the typically anomalous $U(1)$ symmetries are broken.  One would most naturally expect relatively small masses in this case, though, so we will require that no such exotic chiral matter be engineered in the first place.}}.  For this reason, we will require that there be no net hypercharge flux on any of the $\mathbf{10}$ matter curves.

\subsubsection{Precise Statement of Constraints}
\label{subsubsec:preciseconstraints}
We are now ready to finally state the precise constraints that we impose.

\begin{itemize}
\item Turn on a nontrivial hypercharge flux $F_Y$  that is dual in $S_{\rm GUT}$ to a two-cycle, which is a trivial class in $B_3$
\item All three generations of $\mathbf{10}_M$ $(\mathbf{\overline{5}}_M$) must localize on a single matter curve, $\Sigma_{10_M}$ ($\Sigma_{\overline{5}_M}$)
\item The $\mathbf{5}_H$ and $\mathbf{\overline{5}}_H$ fields must localize on distinct matter curves, $\Sigma_{5_H}$ and $\Sigma_{\overline{5}_H}$, which satisfy $F_Y\cdot \Sigma_{5_H}=1$ and $F_Y\cdot \Sigma_{\overline{5}_H}=-1$
\item All $\mathbf{10}$ matter curves, $\Sigma_{\mathbf{10}}^{(i)}$, must satisfy $\Sigma_{\mathbf{10}}^{(i)}\cdot F_Y=0$
\item All MSSM Yukawa couplings, $\mathbf{10}_M\times\mathbf{10}_M\times\mathbf{5}_H$ and $\mathbf{10}_M\times\mathbf{\overline{5}}_M\times\mathbf{\overline{5}}_H$, must be realized
\item The dimension 4 operators $\mathbf{10}_M\times\mathbf{\overline{5}}_M\times\mathbf{\overline{5}}_M$, $\mathbf{10}_M\times\mathbf{\overline{5}}_H\times\mathbf{\overline{5}}_H$ must be forbidden
\end{itemize}

\subsection{Implications of the Constraints}

Let us turn now to the implications of these constraints for the construction of F-theory models, and thier consequences upon the structure of $\mathcal{C}_{\bf 10}$.

\subsubsection{Getting Nontrivial Restrictions on the Superpotential}

The first thing to note is that any nontrivial restriction on superpotential couplings involving $\mathbf{10}$ and $\mathbf{5}$ fields beyond simple $SU(5)$ invariance requires ${\cal{C}}_{10}$ to factorize into at least two components.  To see this, suppose that ${\cal{C}}_{10}$ instead does not factorize.  In this case, the monodromy group $G$ is isomorphic to one of the transitive subgroups of $S_5$ listed in section \ref{monodclass}.  By definition, the $\{\lambda_i\}$ form a unique orbit under the action of $G$ so in order to get any nontrivial structure it is necessary for the set of $\{\lambda_i+\lambda_j\}$, to which the $\mathbf{\overline{5}}$ fields are associated, to comprise more than one orbit.  This rules out all possibilities except $\mathbb{Z}_5$.  Taking $G= \mathbb{Z}_5$ to be generated by $(12345)$, $\{\lambda_i+\lambda_j\}$ splits into two orbits so that we get the following matter curves
\be
\ba
\mathbf{10} & : \quad \{\lambda_1,\lambda_2,\lambda_3,\lambda_4,\lambda_5\} \cr
\mathbf{\overline{5}}^{(1)} &: \quad  \{\lambda_1+\lambda_2,\lambda_2+\lambda_3,\lambda_3+\lambda_4,\lambda_4+\lambda_5,\lambda_5+\lambda_1\} \cr
\mathbf{\overline{5}}^{(2)} &:\quad  \{\lambda_1+\lambda_3,\lambda_2+\lambda_4,\lambda_3+\lambda_5,\lambda_4+\lambda_1,\lambda_5+\lambda_2\} \,.
\ea
\ee
It is now easy to see that all couplings that are allowed by $SU(5)$ invariance can also be obtained from couplings in the "parent" theory that are invariant under the full $U(1)^4$ Cartan of $SU(5)_{\perp}$.

To obtain the structure needed to forbid dimension 4 baryon number violating operators, then, we need that
\begin{center}
\boxed{\textbf{${\cal{C}}_{10}$ factors into at least two components}} \end{center}
 This guarantees that at least one $U(1)$ gauge boson will survive the quotient by $G$.


\subsubsection{Obtaining up-type $\mathbf{10}_M\times\mathbf{10}_M\times\mathbf{5}_H$ Yukawas}
\label{subsubsec:MSSMupYukawa}

Recall that the $\mathbf{10}_M\times\mathbf{10}_M\times\mathbf{5}_H$ coupling originates from points of $E_6$ enhancement where locally
\begin{equation}0=\lambda_i=\lambda_j=\lambda_i+\lambda_j\,,\qquad i\ne j\,.\end{equation}
To realize such a coupling while keeping all three generations of $\mathbf{10}_M$'s  on a single  matter curve, it is necessary that there be a monodromy connecting $\lambda_i$ and $\lambda_j$.  This means that the $\mathbf{10}_M$ matter curve must be contained within a single component, ${\cal{C}}^{(10_M)}$, of ${\cal{C}}_{10}$ of degree at least 2.  Further, the $\mathbf{5}_H$ component must be "contained" in ${\cal{C}}^{(10_M)}$ in the sense that it must arise, in the notation of section \ref{subsubsec:mattcurvC10}, from ${\cal{C}}^{(10_M)}\cap \tau {\cal{C}}^{(10_M)}$.  We can summarize this by saying that
\begin{center}
\boxed{\textbf{$\mathbf{10}_M$ and $\mathbf{5}_H$ matter curves must be contained within a single component of ${\cal{C}}_{10}$}}\end{center}

\subsubsection{Hypercharge Flux and Exotics}\label{subsubsec:hyperexotic}

Let us now turn to the implications of our many conditions related to the hypercharge flux.  To start, note that a generic factorization of ${\cal{C}}_{10}$ takes the form
\begin{equation}{\cal{C}}_{10} = \prod_i {\cal{C}}_{10}^{(i)} = \prod_i \left(\chi^{(i)} V^{m_i} + \ldots)\right)\,,\label{compeqn}\end{equation}
where the classes $[\chi^{(i)}]$ are pulled back from $S_{\rm GUT}$ and the omitted terms $\ldots$ vanish at $U=0$.  It is important to note that the classes of all coefficients that specify this factorization are now determined uniquely in terms of the $[\chi^{(i)}]$ and the first Chern class, $c_1$, of $S_{\rm GUT}$.  The $[\chi^{(i)}]$ are telling us something physical, though -- they are simply the projection of the $\mathbf{10}$ matter curves to $S_{\rm GUT}$.  Our constraint that hypercharge flux restricts trivially to each of these thus amounts to
\begin{equation}F_Y\cdot [\chi^{(i)}] = 0\,.\end{equation}
Triviality of the hypercharge flux in $B_3$ also implies that
\begin{equation}F_Y\cdot c_1=0\,,\end{equation}
which means that the intersection of $F_Y$ with the class of any section appearing as a coefficient in \eqref{compeqn} must vanish.

This has important implications for the $\mathbf{\overline{5}}$ matter curves.  If the monodromy group is as large as possible for a given factorization of ${\cal{C}}_{10}$, namely a product of the relevant symmetric groups, then the $\mathbf{\overline{5}}$ matter curves are uniquely determined by certain polynomials in the coefficients of \eqref{compeqn}.  This means that there is no net hypercharge flux on any $\mathbf{\overline{5}}$ matter curves, making it impossible to properly engineer the Higgs sector!

What we can do for the moment is associate both $H_u$ and $H_d$ to a single matter curve.  From this perspective, they will comprise a vector-like pair that we would generically expect to lift from the spectrum.  This is essentially a statement of the $\mu$ problem, and we will return to the resolution of this issue shortly.  For now, however, we can combine our requirement for $H_u$ and $H_d$ with the result of section \ref{subsubsec:MSSMupYukawa} to obtain
\begin{center}
\boxed{\textbf{$\mathbf{10}_M$, $\mathbf{5}_H$, and $\mathbf{\overline{5}}_H$ matter must all be localized within a single component of ${\cal{C}}_{10}$}}
\end{center}

\subsubsection{Obtaining down-type $\mathbf{10}_M\times\mathbf{\overline{5}}_M\times\mathbf{\overline{5}}_H$ Yukawas}\label{subsubsec:MSSMdownYukawa}

Let us return now to the issue of realizing the MSSM superpotential and recall the origin of the down-type $\mathbf{10}_M\times\mathbf{\overline{5}}_M\times\mathbf{\overline{5}}_H$ Yukawa coupling
\begin{equation}0=\lambda_i+\lambda_j=\lambda_k+\lambda_{\ell}=\lambda_m\,,\qquad i,j,k,\ell,m\text{ distinct}\,.\end{equation}
To obtain such a coupling when $\mathbf{10}_M$ and $\mathbf{\overline{5}}_H$ both localize entirely within a single component ${\cal{C}}^{(10_M)}$ of ${\cal{C}}_{10}$, it is necessary that the degree of ${\cal{C}}^{(10_M)}$ be at least 3.  In other words,
\begin{center}
\boxed{
\textbf{$\mathbf{10}_M$, $\mathbf{5}_H$, and $\mathbf{\overline{5}}_H$ must localize in a single component of ${\cal{C}}_{10}$ } 
 \textbf{of degree at least 3}}
\end{center}

\subsubsection{Three Options$\ldots$so far}\label{subsubsec:threeoptions}

The factorizations of ${\cal{C}}_{\mathbf{10}}$ that are consistent with everything we have said thus far are
\begin{itemize}
\item ${\cal{C}}_{\mathbf{10}} = {\cal{C}}_{\mathbf{10}}^{(3)}+{\cal{C}}_{\mathbf{10}}^{(2)}$
\item ${\cal{C}}_{\mathbf{10}} = {\cal{C}}_{\mathbf{10}}^{(3)}+{\cal{C}}_{\mathbf{10}}^{(1),1}+{\cal{C}}_{\mathbf{10}}^{(1),2}$
\item ${\cal{C}}_{\mathbf{10}} = {\cal{C}}_{\mathbf{10}}^{(4)}+{\cal{C}}_{\mathbf{10}}^{(1)}$
\end{itemize}
where we have indicated the degree of each factor.  In each case, the matter fields arise from the following components of ${\cal{C}}_{\mathbf{10}}$ modulo the necessary subtractions for isolating ${\bf 5}$ matter curves, explained in section \ref{subsubsec:mattcurvC10}

\begin{equation}\begin{array}{c|c|c|c}\text{Factorization} & \mathbf{10}_M & \mathbf{5}_H + \mathbf{\overline{5}}_H & \mathbf{\overline{5}}_M \\ \hline
{\cal{C}}_{\mathbf{10}}^{(3)}+{\cal{C}}_{\mathbf{10}}^{(2)} & (U=0)\cap {\cal{C}}_{\mathbf{10}}^{(3)} & {\cal{C}}_{\mathbf{10}}^{(3)}\cap \tau {\cal{C}}_{\mathbf{10}}^{(3)} & {\cal{C}}_{\mathbf{10}}^{(2)}\cap\tau {\cal{C}}_{\mathbf{10}}^{(2)} \\
{\cal{C}}_{\mathbf{10}}^{(3)}+{\cal{C}}_{\mathbf{10}}^{(1),1}+{\cal{C}}_{\mathbf{10}}^{(1),2} & (U=0)\cap {\cal{C}}_{\mathbf{10}}^{(3)} &{\cal{C}}_{\mathbf{10}}^{(3)}\cap \tau{\cal{C}}_{\mathbf{10}}^{(3)} & {\cal{C}}_{\mathbf{10}}^{(1),1}\cap \tau{\cal{C}}_{\mathbf{10}}^{(1),2} + {\cal{C}}_{\mathbf{10}}^{(1),2}\cap \tau{\cal{C}}_{\mathbf{10}}^{(1),1} \\
{\cal{C}}_{\mathbf{10}} = {\cal{C}}_{\mathbf{10}}^{(4)}+{\cal{C}}_{\mathbf{10}}^{(1)} & (U=0)\cap {\cal{C}}_{\mathbf{10}}^{(4)} & {\cal{C}}_{\mathbf{10}}^{(4)}\cap \tau {\cal{C}}_{\mathbf{10}}^{(4)} & {\cal{C}}_{\mathbf{10}}^{(4)}\cap \tau{\cal{C}}_{\mathbf{10}}^{(1)} + {\cal{C}}_{\mathbf{10}}^{(1)}\cap \tau{\cal{C}}_{\mathbf{10}}^{(4)} \\
\end{array}\end{equation}
Further, in each case we realize one anomalous $U(1)$ gauge symmetry for controlling the superpotential under which the MSSM fields are charged{\footnote{The case ${\cal{C}}_{\mathbf{10}}^{(3)}+{\cal{C}}_{\mathbf{10}}^{(1),1}+{\cal{C}}_{\mathbf{10}}^{(1),2}$ realizes a second anomalous $U(1)$ gauge symmetry but it does not couple to any of the MSSM fields.}}.  In fact, the $U(1)$ charges of MSSM matter fields are identical in each case:{\footnote{This occurs because the charges are completely fixed by the fact that all MSSM Yukawas are realized combined with our requirement that the Higgs multiplets, $\mathbf{5}_H$ and $\mathbf{\overline{5}}_H$, arise from the same component (and hence have opposite charges)}}
\begin{equation}\begin{array}{c|c}\text{Field} & $U(1)$ \\ \hline
\mathbf{10}_M & 1 \\
\mathbf{\overline{5}}_M & -3 \\
\mathbf{5}_H & -2 \\
\mathbf{\overline{5}}_H & 2 \\
\end{array}\end{equation}
Quite nicely, this $U(1)$, which is a combination of $U(1)_Y$ and $U(1)_{B-L}$, is sufficient to forbid all baryon number violating dimension 4 operators.

\subsubsection{The $\mu$ Problem and $U(1)_{PQ}$}
\label{U1PQmu}

While we have identified three candidate factorizations of ${\cal{C}}_{10}$ that satisfy most of our desired constraints, recall from the discussion of section \ref{subsubsec:hyperexotic} that we have a potentially serious problem with the Higgs sector.  This is because we were forced to require $H_u$ and $H_d$ to localize within the same component of ${\cal{C}}_{10}$.  If the monodromy group is the generic one for such a situation, namely a product of symmetric groups, then $H_u$ and $H_d$ must come from the same matter curve.  This matter curve has no net hypercharge flux so what we require to get the Higgs sector right, then, is a nongeneric situation in which a vector-like pair on a single matter curve remains massless in the absence of any symmetry that guarantees it.  This is just a restatement of the standard $\mu$ problem, which we would like to avoid.

To resolve this issue, it is necessary to refine the geometry so that the monodromy group $G$ is not a product of symmetric groups but rather a suitable subgroup thereof.  However, it is important that $G$ not be so small of a subgroup that ${\cal{C}}_{10}$ factors beyond the options presented in section \ref{subsubsec:threeoptions}.
As we saw at the end of section \ref{subsubsec:imprint}, such a refinement is not possible unless the number of components of ${\cal{C}}_{10}$ is 4 or 5.  If all of the components of ${\cal{C}}_{10}$ are linear or quadratic, $G$ is uniquely determined by the factorization of ${\cal{C}}_{10}$.  Further, when ${\cal{C}}_{10}$ has a cubic piece the only transitive subgroup of $S_3$ is $A_3$, which does not lead to any further refinement of the $\mathbf{5}$ matter curves of the type that we need for our Higgs sector.


For these reasons, we are forced to consider the 4+1 factorization,
\begin{equation}{\cal{C}}_{10} = {\cal{C}}_{10}^{(4)}+{\cal{C}}_{10}^{(1)}\,.\label{4p1factorization}\end{equation}
In this case, we need $G$ to be a proper transitive subgroup of $S_4$ whose action on the set $\{\lambda_i+\lambda_j\}$, $1\le i\ne j\le 4$, decomposes into at least two distinct orbits. 
 From the list of possibilities of section \ref{monodclass}, it is easy to see that only three have this property.  These are
\begin{equation}G=\mathbb{Z}_4,\text{ }D_4,\text{ or }V\,.\end{equation}

\subsection{The cases $G=\mathbb{Z}_4$, $D_4$, and $V$}

We now describe the structure of theories with monodromy groups $G=\mathbb{Z}_4$, $D_4$, and $V$ in detail in order to explicitly see how all of our conditions are satisfied.

We start with the case $G=\mathbb{Z}_4$ and consider, for illustration, the specific $\mathbb{Z}_4$ subgroup of $S_5$ generated by $(1234)$.  We obtain several matter curves associated with the following $\lambda_i$
\begin{equation}
\begin{split}\label{Z4Mat}
\mathbf{10}^{(1)} : &\quad \{\lambda_1,\lambda_2,\lambda_3,\lambda_4\} \\
\mathbf{10}^{(2)} : &\quad \{\lambda_5\} \\
\mathbf{5}^{(1)} : &\quad \{\lambda_1+\lambda_2,\lambda_2+\lambda_3,\lambda_3+\lambda_4,\lambda_4+\lambda_1\} \\
\mathbf{5}^{(2)}: &\quad \{\lambda_1+\lambda_3,\lambda_2+\lambda_4\} \\
\mathbf{5}^{(3)}: &\quad \{\lambda_1+\lambda_5,\lambda_2+\lambda_5,\lambda_3+\lambda_5,\lambda_4+\lambda_5\} \\
\end{split}\end{equation}
From this, it is easy to see that if we identify $\mathbf{10}_M\sim \mathbf{10}^{(1)}$, $\mathbf{\overline{5}}_M\sim \mathbf{\overline{5}}^{(3)}$, and $\mathbf{5}_H,\mathbf{\overline{5}}_H$ with either $\mathbf{5}^{(1)}$,$\mathbf{\overline{5}}^{(2)}$ or $\mathbf{5}^{(1)}$,$\mathbf{\overline{5}}^{(2)}$, then all of the MSSM couplings are present
\begin{equation}W_{MSSM}\sim \mathbf{10}_M\times\mathbf{10}_M\times\mathbf{5}_H + \mathbf{10}_M\times\mathbf{\overline{5}}_M\times\mathbf{\overline{5}}_H\,,\label{Z4105}\end{equation}
while a bare $\mu$ term is forbidden
\begin{equation}W_{\mu}\sim \text{\sout{$\mathbf{5}_H\times\mathbf{\overline{5}}_H$}}\,.\end{equation}
Further, hypercharge flux restricts trivially on the  ${\bf 10}$ matter curve, and thus we can avoid getting charged exotics. 
At the level of renormalizable couplings, the effective action respects an accidental $U(1)_{PQ}$ global symmetry
\begin{equation}\begin{array}{c|c}\text{Field} & U(1)_{PQ} \\ \hline
\mathbf{10}_M & -1 \\
\mathbf{\overline{5}}_M & -1 \\
\mathbf{5}_H & 2 \\
\mathbf{\overline{5}}_H & 2
\label{U1PQ}\end{array}\end{equation}
that may nonetheless be broken by certain nonrenormalizable operators.

From the perspective of couplings involving $\mathbf{10}$, $\mathbf{5}$ fields and their conjugates, the $D_4$ case is identical to this.  As an explicit example, we can consider the $D_4$ subgroup generated by $(1234)$ and $(13)$.  In this case, the breakdown of $\mathbf{10}$ and $\mathbf{5}$ matter curves is identical to \eqref{Z4Mat}.

Turning now to the case $G=V=\{1,(12)(34),(13)(24),(14)(23)\}$, the structure of $\mathbf{5}$ matter curves is slightly more refined.  In particular, the $\mathbf{5}^{(1)}$ matter curve splits in two, leading to
\begin{equation}
\begin{split}
\mathbf{10}^{(1)} : &\quad \{\lambda_1,\lambda_2,\lambda_3,\lambda_4\} \\
\mathbf{10}^{(2)} : &\quad \{\lambda_5\} \\
\mathbf{5}^{(1)} : &\quad \{\lambda_1+\lambda_2,\lambda_3+\lambda_4\} \\
\mathbf{5}^{(1)'} : &\quad \{\lambda_1+\lambda_4,\lambda_2+\lambda_3\} \\
\mathbf{5}^{(2)}: &\quad \{\lambda_1+\lambda_3,\lambda_2+\lambda_4\} \\
\mathbf{5}^{(3)}: &\quad \{\lambda_1+\lambda_5,\lambda_2+\lambda_5,\lambda_3+\lambda_5,\lambda_4+\lambda_5\} \\
\end{split}\end{equation}

Nevertheless, if we identify $\mathbf{10}_M\sim \mathbf{10}^{(1)}$, $\mathbf{\overline{5}}_M\sim \mathbf{\overline{5}}^{(3)}$, and $\mathbf{5}_H,\mathbf{\overline{5}}_H$ with some pair $\mathbf{5}^{(a)},\mathbf{\overline{5}}^{(b)}$ associated to distinct matter curves then we again realize the MSSM superpotential while forbidding dimension 4 proton decay operators and a bare $\mu$ term.  Because of this, $U(1)_{PQ}$ \eqref{U1PQ} again arises as an accidental global symmetry of the action at the level of renormalizble couplings.

\subsection{Implications for SUSY-Breaking and Neutrino Physics}

We would like to know if one of the restricted set of scenarios that have been singled out thus far can even in principle accommodate any of the successes of local models regarding gauge mediation and the generation of neutrino masses.  The answer to this is not immediately obvious because the scenarios that seem to emerge from our constraints have not explicitly appeared in the literature thus far.

The crucial difference between the structures that we find and those that have been studied in the local context is the presence or absence of a gauged $U(1)_{PQ}$ symmetry responsible for forbidding a bare $\mu$ term.  For us, only one gauged $U(1)$ was allowed and this was a combination of $U(1)_{B-L}$ and $U(1)_Y$.  Any $U(1)_{PQ}$ arose as an accidental global symmetry of the renormalizable Lagrangian.  On the other hand, studies of local models often take a gauged $U(1)_{PQ}$ as an important starting point \cite{Ibe:2007km,Marsano:2008jq,Heckman:2008qt}.  What we would like to ask now is how crucial this gauge symmetry actually is.

To study this issue further, it is necessary to go beyond a simple analysis of the $\mathbf{10}$, $\mathbf{5}$ matter curves and include also the matter curves on which GUT singlets, $\mathbf{1}$, localize.  These are identified with weights $\lambda_i-\lambda_j$ so it is easy to work out their structure in the three cases of interest.  Let us start again with the case $G=\mathbb{Z}_4$ since this will illustrate the most important points.  There, in addition to the $\mathbf{10}$ and $\mathbf{5}$ matter curves
\begin{equation}
\begin{split}
\mathbf{10}_M : &\quad \{\lambda_1,\lambda_2,\lambda_3,\lambda_4\} \\
\mathbf{10}^{(2)} : &\quad \{\lambda_5\} \\
\mathbf{5}^{(1)} : &\quad \{\lambda_1+\lambda_2,\lambda_2+\lambda_3,\lambda_3+\lambda_4,\lambda_4+\lambda_1\} \\
\mathbf{5}^{(2)}: &\quad \{\lambda_1+\lambda_3,\lambda_2+\lambda_4\} \\
\mathbf{\overline{5}}_M: &\quad \{\lambda_1+\lambda_5,\lambda_2+\lambda_5,\lambda_3+\lambda_5,\lambda_4+\lambda_5\} \\
\end{split}\end{equation}
where $\mathbf{5}_H,\mathbf{\overline{5}}_H\sim \mathbf{5}^{(i)},\mathbf{\overline{5}}^{(j)}$ with $i\ne j$, we have also the singlet matter curves
\begin{equation}\begin{split}
\mathbf{1}^{(1)}: &\quad \{\lambda_1-\lambda_2,\lambda_2-\lambda_3,\lambda_3-\lambda_4,\lambda_4-\lambda_1\} \\
\mathbf{1}^{(1)\prime} &\quad \{-(\lambda_1-\lambda_2),-(\lambda_2-\lambda_3),-(\lambda_3-\lambda_4),-(\lambda_4-\lambda_1)\} \\
\mathbf{1}^{(2)}: &\quad \{\lambda_1-\lambda_3,\lambda_2-\lambda_4,-(\lambda_1-\lambda_3),-(\lambda_2-\lambda_4)\} \\
\mathbf{1}^{(3)}: &\quad \{\lambda_1-\lambda_5,\lambda_2-\lambda_5,\lambda_3-\lambda_5,\lambda_4-\lambda_5\} \\
\mathbf{1}^{(3)\prime}: &\quad \{-(\lambda_1-\lambda_5),-(\lambda_2-\lambda_5),-(\lambda_3-\lambda_5),-(\lambda_4-\lambda_5)\}
\end{split}\end{equation}
In the supersymmetry breaking scenarios of \cite{Ibe:2007km,Marsano:2008jq,Heckman:2008qt}, the $\mu$ term is generated via a Giudice-Masiero type mechanism involving a coupling of the form
\begin{equation}\int\,d^4\theta\,\frac{X^{\dag} H\overline{H}}{\Lambda}\,.\end{equation}
Here, $X$ is a singlet field whose $F$-component expectation value is responsible for breaking supersymmetry.  When $(\mathbf{5}_H,\mathbf{\overline{5}}_H)\sim (\mathbf{5}^{(1)},\mathbf{\overline{5}}^{(2)})$, this coupling arises if we identify $X\sim \mathbf{1}^{(1)}$.  If instead, we have $(\mathbf{5}_H,\mathbf{\overline{5}}_H)\sim (\mathbf{5}^{(2)},\mathbf{\overline{5}}^{(1)})$ then it is necessary to identify $X\sim \mathbf{1}^{(1)\prime}$.  Further, in both cases a coupling of the form
\begin{equation}\int\,d^4\theta\,\frac{|X|^2 H\overline{H}}{\Lambda^2}\,,\label{Bmuop}\end{equation}
is forbidden.  It was this coupling that could generate a $B_{\mu}$ term and its absence assures that the solution to the $\mu/B_{\mu}$ problem in \cite{Ibe:2007km,Marsano:2008jq,Heckman:2008qt} is retained.  Before moving on to neutrinos, we should note that the spurion field $X$ is no longer charged under a gauged $U(1)$ symmetry.  This means that models of this sort do not lead to a "PQ-deformed" gauge mediation scenario \cite{Heckman:2008qt} but instead a scenario closer in spirit to \cite{Ibe:2007km}.  Further, we do not immediately know the consequences of this lack of a gauged $U(1)_{PQ}$ on the ability of stringy instantons to trigger supersymmetry-breaking along the lines of \cite{Heckman:2008es,Marsano:2008py}.  It would be interesting to address this further.

To discuss neutrino physics, we focus on the two principal scenarios described in \cite{Bouchard:2009bu,Heckman:2009mn}.  The first is a Majorana scenario in which KK modes play the role of a tower of "right handed neutrinos" and, when integrated out, generate an operator of the form
\begin{equation}\int\,d^2\theta\,\frac{(H_u L)^2}{\Lambda^2}\,.\end{equation}
This descends from the $\left(\mathbf{5}_H\times\mathbf{\overline{5}}_M\right)^2$ and is expressly forbidden by our gauged $U(1)_{B-L}$. This is not to say that Majorana neutrinos masses cannot be accommodated at all, but rather their generation must, as usual, be associated with the breaking of $U(1)_{B-L}$. 

Turning now to the Dirac scenario of \cite{Bouchard:2009bu}, we recall that it is based on an operator of the form
\begin{equation}\int\,d^4\theta\, \frac{H_d^{\dag} L N_R}{\Lambda}\,,\label{diracscenario}\end{equation}
where $N_R$ is a right-handed neutrino.  
Because $H_d$ picks up an $F$-component expectation value from the $\mu$ term proportional to $[H_d]_{\theta^2}\sim \mu <H_u>$, this leads to a Dirac neutrino mass with suppression factor $\mu/\Lambda$.  Of the possible candidates for $N_R$, only $\mathbf{1}^{(3)}$ is suitable.  Once this choice is made, we see that the desired operator is allowed regardless of whether we choose $\mathbf{5}_H,\mathbf{\overline{5}}_H\sim \mathbf{5}^{(1)},\mathbf{\overline{5}}^{(2)}$ or $\mathbf{5}_H,\mathbf{\overline{5}}_H\sim \mathbf{5}^{(2)},\mathbf{\overline{5}}^{(1)}$.  Note, however, that a standard Dirac mass operator is also allowed
\begin{equation}
\int\,d^2\theta\, H_u L N_R\,,
\label{baredirac}\end{equation}
which arise with $\mathcal{O}(1)$ coefficient from the point of $E_8$ enhancement, that is required to minimize flavor mixing \cite{Bouchard:2009bu, Heckman:2009mn}. 
The beautiful suppression of neutrino masses from \eqref{diracscenario} is unfortunately lost because we have nothing to prevent a "bare" Dirac mass that overwhelms it.  This problem is related to  the lack of a gauged $U(1)_{PQ}$ symmetry.  Without it, \eqref{diracscenario} and \eqref{baredirac} are on an equal footing as far as their charges under $U(1)$ gauge symmetries are concerned.   We can make an assignment of $U(1)_{PQ}$ charges that allows \eqref{baredirac}, reflecting the fact that it arises as a global symmetry of the renormalizable Lagrangian, but nothing \emph{a priori} prevents the generation of higher dimension operators such as \eqref{baredirac} Êthat violate it.  In the case of the $B_{\mu}$ generating operator \eqref{Bmuop}, we were lucky enough that it was forbidden anyway.  With neutrinos, we have not been so fortunate.

It is easy to verify that considering $G=D_4$ or $G=V$ will not improve the situation significantly.  The existence of a gauged $U(1)_{B-L}$ doomed the KK Majorana scenario while the lack of a gauged $U(1)_{PQ}$ doomed the Dirac scenario.

To summarize, then, our restricted set of scenarios shows some promising signs of being able to accommodate  the general gauge mediation framework that has emerged from studies of local models \cite{Marsano:2008jq, Heckman:2008qt}.  However, there seem to be serious problems with neutrinos.


\subsubsection{Relaxing the Constraints?}
\label{subsubsec:relax}

Because of the tension with the successful neutrino scenarios of \cite{Bouchard:2009bu,Heckman:2009mn}, it would be interesting to determine whether any of the constraints in section \ref{subsubsec:preciseconstraints} can be relaxed.  In this subsection, we make some preliminary comments about this issue.

The most stringent constraint is the requirement that $F_Y\cdot \Sigma_{10}^{(i)}=0$ for all $\mathbf{10}$ matter curves, $\Sigma_{10}^{(i)}$.  This is what forced us to realize both $\mathbf{5}_H$ and $\mathbf{\overline{5}}_H$ on the same component of ${\cal{C}}_{10}$ and hence it is responsible for $U(1)_{B-L}$, rather than $U(1)_{PQ}$, emerging as the gauged $U(1)$ that forbids dimension 4 proton decay operators.  What if we relaxed this constraint?

If we allow $F_Y\cdot \Sigma_{10}^{(i)}\ne 0$ on some $\mathbf{10}$ matter curves then we are guaranteed to get charged exotics that do not comprise a full GUT multiplet.  From the $SU(5)$ point of view they will come in vector-like pairs but they will be distinguished by their charges under $U(1)$ gauge symmetries that remain after the quotient.  For simplicity, let us suppose that nontrivial hypercharge flux threads only two $\mathbf{10}$ matter curves, $\Sigma_{10}^{(1)}$ and $\Sigma_{10}^{(2)}$.  The corresponding exotics cannot couple to one another directly because the term $\mathbf{10}^{(1)}\times \mathbf{\overline{10}}^{(2)}$ is forbidden.  However, they can couple through a singlet field via  operators that descend from terms like
\begin{equation}\mathbf{10}^{(1)}\times \mathbf{\overline{10}}^{(2)}\times \mathbf{1}\,,\end{equation}
which like all other couplings originates from ${\bf 248}^3$ of $E_8$.
If some dynamics causes this singlet to pick up a nonzero bosonic expectation value then the unwanted exotics can pick up a nonzero mass.

Quite remarkably, something similar already happens in the gauge mediation scenarios required by $E_8$ unification in \cite{Heckman:2009mn}.  There, some extra $\mathbf{10}$'s localize on the $\mathbf{10}_M$ matter curve while some $\mathbf{\overline{10}}$'s localize on a different $\mathbf{10}$ matter curve.  These couple to the spurion field $X$ that triggers supersymmetry breaking via
\begin{equation}X\times \mathbf{10}\times \mathbf{\overline{10}}\,,\end{equation}
and hence play the role of gauge messengers.  When $X$ picks up a nonzero expectation value, they get a mass proportional to $\langle X\rangle$ which, for typical F-theory GUT scenarios, is only a few orders of magnitude below $M_{GUT}$.

A possible remedy for our present troubles\footnote{
Another possibility is to include higher order terms in the expansion of $f$ and $g$ in powers of $z$ in section \ref{sec:Fourfold}. This could possibly provide us with more freedom.  }, then, is to allow hypercharge flux to thread the matter curves on which the gauge messengers localize in the models of \cite{Heckman:2009mn}.  The dynamics of supersymmetry breaking will push the exotics, which are now playing the role of gauge messengers, to a relatively high scale, namely the messenger scale\footnote{Recall, that in the scenarios of \cite{Marsano:2008jq, Heckman:2008qt} the expectation value is of order $10^{12} - 10^{14}$ GeV.}.  If we do this, then it is possible to realize all of the other constraints of section \ref{subsubsec:preciseconstraints} with the monodromy groups of \cite{Heckman:2009mn}.  

\subsubsection{Incomplete GUT Multiplets and Unification}
\label{subsubsec:unification}

What about the effect on gauge coupling unification?  While these incomplete GUT multiplets will prevent unification, it is important to note that unification is already disrupted at $M_{GUT}$ in these models \cite{Blumenhagen:2008aw}.  This is because the internal hypercharge flux modifies the $SU(3)$, $SU(2)$, and $U(1)_Y$ gauge kinetic terms at $M_{GUT}$ through the Chern-Simons coupling
\begin{equation}S_{CS}\sim \int_{\mathbb{R}^{3,1}\times S_{\rm GUT}}\,C_0\wedge \text{tr}\left(F^4\right)\,.\end{equation}
As demonstrated in \cite{Blumenhagen:2008aw}, this splitting means that the strong condition of gauge coupling unification is replaced by the slightly weaker one that  the various MSSM gauge couplings at the GUT scale, $\alpha_3^{-1}(M_{GUT})$, $\alpha_2^{-1}(M_{GUT})$, and $\alpha_1^{-1}(M_{GUT})$, must satisfy
\begin{equation}\alpha_1^{-1}(M_{GUT})-\frac{3}{5}\alpha_2^{-1}(M_{GUT}) - \frac{2}{5}\alpha_3^{-1}(M_{GUT})=0\,.\label{couplingconstraint}\end{equation}
Further, we expect that the couplings will \emph{not} precisely unify, a rather striking claim given that we know that the MSSM matter spectrum is fairly consistent with unification to within a few percent.
To account for this, it is in fact \emph{necessary} to introduce some matter fields at a high scale $M$ that form incomplete GUT multiplets and whose net contributions, $\delta b_i$, to the various MSSM $\beta$ function coefficients, $b_i$, satisfy
\begin{equation}\delta b_1 - \frac{3}{5}\delta b_2 - \frac{2}{5}\delta b_3=0\,.\label{blumenhagenconstraint}\end{equation}
In this case, the "apparent" unification at scales below $M$ arises due to a cancellation between the effects of the hypercharge flux and the new massive fields.  

The easiest way to get such incomplete GUT multiplets is to engineer them on matter curves threaded by nontrivial hypercharge flux, just as we were forced to do in order to make favorable neutrino physics possible.  Moreover, it is easy to verify directly that any incomplete multiplets that arise in this way, either on $\mathbf{10}$ or $\mathbf{5}$ matter curves with nontrivial hypercharge flux, automatically satisfy \eqref{blumenhagenconstraint}.  This is reviewed briefly in Appendix \ref{app:mattercurves}.  

In the end, we reach the rather surprising conclusion that favorable neutrino physics forces the introduction of incomplete GUT multiplets which, in turn, can account for the fact that the gauge couplings do not really unify.  Solving one problem provides the solution to a second for free.  The simplest way to incorporate these incomplete GUT multiplets is to realize them as messenger fields, at which point we expect them to provide fairly distinct experimental signatures.  With LHC data on the horizon, this would be very interesting to investigate further.

\section{Factorization of the Spectral Surface of Type $4+1$}
\label{sec:factorization}

We now turn our attention to scenarios in which the spectral surface factors into a linear and a quartic piece, $\mathcal{C}_{\mathbf{10}}^{(1)} + \mathcal{C}_{\mathbf{10}}^{(4)}$.  We hope to address scenarios of the type discussed in section \ref{subsubsec:relax} in the future \cite{futurework}.

In this section we discuss the properties of the factorized spectral surface as well as the construction of fluxes. In the next section we will then give examples of three-generation $SU(5)$ GUT models.  These will have the generic $S_4$ monodromy group but we will comment on how to obtain instead one of the subgroups $V$, $\mathbb{Z}_4$, or $D_4$ in section \ref{sec:mu}.

\subsection{Factorized Spectral Surface}
As we discussed in Section 2, the fundamental spectral surface ${\cC}_{\mathbf{10}}$ is a useful object for studying the monodromy group $G$ that controls the structure of matter curves and the superpotential.
In general,
\begin{equation}
\mathcal{C}_{\mathbf{10}}\,: \qquad  b_0\prod_{i=1}^5(s+\lambda_i) = b_0 s^5 + b_2s^3 + b_3s^2+b_4s + b_5  =0\,.
\end{equation}
is irreducible so that there is very little structure.  In this case, all $\mathbf{10}$ ($\mathbf{\overline{5}}$) multiplets that descend from the $E_8$ adjoint are identified under monodromies and the superpotential contains all allowed $SU(5)$-invariant couplings.

As we have seen, to get enough structure that the constraints of section \ref{subsubsec:preciseconstraints} are satisfied, it is necessary for the spectral surface to factor into quartic and linear pieces as


\begin{equation}
\mathcal{C}_{\bf 10}:\qquad  (a_0s^4+a_1s^3+a_2s^2+a_3s+a_4)(d_0s+ d_1) =0\,,
\end{equation}
with
\begin{equation}b_1=a_0d_1+d_0a_1=0\,.
\label{b1cond}\end{equation}

For generic $a_m,d_n$, the monodromy group here is $S_4$ and the sheets, $\lambda_i$, form two distinct orbits $\{\lambda_1,\lambda_2,\lambda_3,\lambda_4\}$ and $\{\lambda_5\}$.  We get two sorts of $\mathbf{10}$ and $\mathbf{\overline{5}}$ matter curves whose charges under the single remaining gauged $U(1)$ symmetry are given in the following table
\begin{equation}
\begin{array}{c|c|c}\text{Matter} & \text{Type of Matter Curve} & U(1)\text{ Charge} \\ \hline
\mathbf{10}_M & \lambda_i=0 & 1 \\
\mathbf{\overline{5}}_H & \lambda_i+\lambda_j=0 & 2 \\
\mathbf{\overline{5}}_M & \lambda_i+\lambda_5=0 & -3 \\
\mathbf{10}_{\text{other}} & \lambda_5=0 & -4
\end{array}
\end{equation}
Our choice of notation here is intentional as the cubic superpotential couplings with this assignment are precisely
\begin{equation}\mathbf{10}_M\times\mathbf{10}_M\times\mathbf{5}_H + \mathbf{10}_M\times\mathbf{\overline{5}}_M\times\mathbf{\overline{5}}_H+\mathbf{10}_{\text{other}}\times\mathbf{\overline{5}}_H\times\mathbf{\overline{5}}_H\end{equation}
This is of course precisely the MSSM superpotential with no additional proton decay operators provided we localize $\mathbf{10}_M$, $\mathbf{\overline{5}}_M$, $\mathbf{5}_H$, and $\mathbf{\overline{5}}_H$ on the indicated matter curves while avoiding the generation of any $\mathbf{10}_{\text{other}}$ zero modes.  One easy way quickly remove the $\mathbf{10}_{\text{other}}$'s is as follows.  Quite intentionally, we have not specified the bundles of which the objects $a_m,d_n$ are sections.  A convenient choice is to take $d_1$ to be a section of the trivial bundle, ${\cal{O}}$, on $X$ so that, in particular, $d_1$ is a number that can effectively be replaced by 1.  This effectively removes the $\mathbf{10}_{\text{other}}$ matter curve, whose defining equation is $d_1=0$, and fixes the remaining classes as
\begin{equation}\begin{array}{c|c}\text{Section} & \text{Divisor Class} \\ \hline
U & \sigma \\
V & \sigma_{\infty}=\sigma+\pi^{\ast}(c_1) \\
d_0 & \pi^{\ast}(c_1) \\
d_1 & {\cal{O}} \\
a_m & \pi^{\ast}\left(\eta - (m+1)c_1\right)
\end{array}\end{equation}
It also means that, in order to satisfy \eqref{b1cond}, $a_0$ must be a product
\begin{equation}\label{b1zero}
a_0 = -a_1d_0\,.\end{equation}
To make all of this clear, we will write the equation for ${\cal{C}}_{\mathbf{10}}$ as
\begin{equation}{\cal{C}}_{\mathbf{10}}:\qquad \left(a_1U^3\left[V-d_0U\right]+a_2U^2V^2+a_3UV^3+a_4V^4\right)\left(V+d_0U\right)\,.\end{equation}
The relation to the coefficients $b_m$ is now given by
\be
\label{param}
\ba
b_5 & = a_4 \cr
b_4 & = a_3 + a_4 d_0 \cr
b_3 & = a_2 + a_3 d_0 \cr
b_2 & = a_1 + a_2 d_0 \cr
b_0 & = - a_1 d_0^2 \,,
\ea
\ee
where we used the constraint (\ref{b1zero}) that arises from $b_1=0$.


\subsection{Matter Curves}

We can analyze the various matter curves in more detail, by recalling that they  originate from the intersection of ${\cal{C}}_{\mathbf{10}}$ with its image under $V\rightarrow -V$.  In this case, that image is
\begin{equation}
\tau{\cal{C}_{\mathbf{10}}}:\quad-\left(-a_1U^3\left[V+d_0U\right]+a_2U^2V^2-a_3UV^3+a_4V^4\right)\left(V-d_0U\right) \,.
\end{equation}
In appendix \ref{app:Matter41} we study the decomposition of ${\cal{C}_{\mathbf{10}}}\cap \tau{\cal{C}_{\mathbf{10}}}$ into the three pieces described in section \ref{subsubsec:mattcurvC10}.
We now summarize, how the three components of ${\cal{C}_{\mathbf{10}}}\cap\tau{\cal{C}_{\mathbf{10}}}$ split among the factors ${\cal{C}}_{\bf 10}^i\cap\tau{\cal{C}}_{\bf 10}^j$.  First we recall the classes of ${\cal{C}}_{\mathbf{10}}$ and its two components inside $X$,
\begin{equation}
{\cal{C}_{\mathbf{10}}}: \  5\sigma+\pi^{\ast}\eta\,,\qquad
{\cal{C}}_{\mathbf{10}}^{(1)}: \  \sigma_{\infty} \,, \qquad
{\cal{C}}_{\mathbf{10}}^{(4)}: \   4\sigma+\pi^{\ast}(\eta-c_1) \,.
\end{equation}
Then the various factors of ${\cal{C}}_{\mathbf{10}}\cap \tau {\cal{C}}_{\mathbf{10}}$ split as follows:
{\small
\begin{equation}\begin{array}{c||c|c|c||c}
\text{Type} & {\cal{C}}_{\bf 10}^{(1)}\cap\tau{\cal{C}}_{\bf 10}^{(1)} & {\cal{C}}_{\bf 10}^{(1)}\cap \tau{\cal{C}}_{\bf 10}^{(4)}+ {\cal{C}}_{\bf 10}^{(4)}\cap \tau{\cal{C}}_{\bf 10}^{(1)} & {\cal{C}}_{\bf 10}^{(4)}\cap\tau{\cal{C}}_{\bf 10}^{(4)} & \text{Total in }{\cal{C}_{\bf 10}}\cap \tau{\cal{C}_{\bf 10}} \\ \hline
\mathbf{10}\text{ M.C.} & \cdot & \cdot & \sigma\cdot\pi^{\ast}(\eta-5c_1) & \sigma\cdot\pi^{\ast}(\eta-5c_1)\\ \hline
\mathbf{5}\text{ M.C.} & \cdot & 2\sigma\cdot\pi^{\ast}(\eta-3c_1) & 2\sigma\cdot \pi^{\ast}(2\eta-7c_1) & 2\sigma\cdot\pi^{\ast}(3\eta-10c_1) \\
& \cdot & + 2\pi^{\ast}(c_1)\cdot \pi^{\ast}(\eta-3c_1) & + \pi^{\ast}(\eta-2c_1)\cdot\pi^{\ast}(\eta-3c_1) & + \pi^{\ast}(\eta)\cdot \pi^{\ast}(\eta-3c_1) \\ \hline
 \cap \text{ at }\infty & \sigma_{\infty}\cdot\pi^{\ast}c_1 & 4\sigma_{\infty}\cdot \pi^{\ast}c_1 & \sigma_{\infty}\cdot\pi^{\ast}(3\eta-5c_1) & 3\sigma_{\infty}\cdot \pi^{\ast}\eta \\ \hline \hline
\text{Total} & \sigma_{\infty}\cdot \pi^{\ast}c_1 & 2\times \left(\sigma_{\infty}\cdot \pi^{\ast}(\eta-c_1)\right)
& 8\sigma\cdot \pi^{\ast}(\eta-3c_1)& (5\sigma+\pi^{\ast}\eta)^2 \\
& & & +\pi^{\ast}(\eta-c_1)^2  &
\end{array}\end{equation}}
Here we list both the contributions to the ${\bf 10}$ and ${\bf 5}$ matter curves, as well as the intersection at infinity.
We see that the "Total" column is precisely the net class of the various components of ${\cal{C}}\cap \tau{\cal{C}}$.



\subsection{Yukawa structure}

On general grounds, we expect this factorization to produce two distinct $\mathbf{\overline{5}}$ matter curves and intersections that produce precisely the MSSM superpotential couplings and nothing more.  We can see this explicitly as follows.  
 
First, recall that the defining equation for the $\mathbf{10}$ matter curve is
\begin{equation}0=b_5=a_4\end{equation}
As expected, setting $d_1=1$ has left us with a single $\mathbf{10}$ matter curve.  Now, note that the defining equation for the $\mathbf{\overline{5}}$ matter curve is given by
\be 
0= P  = b_0b_5^2-b_2b_3b_5+b_3^2b_4= \left(a_3 \left(a_2+a_3 d_0\right)-a_1 a_4\right)
\left(a_2+d_0 \left(a_3+a_4 d_0\right)\right)   \,.
\ee 
which is automatically factored.
The two components are
denoted by $P_{H, M}$ \be\label{intermsofa} P_H =  \left(a_3
\left(a_2+a_3 d_0\right)-a_1 a_4\right)\,,\qquad P_M= \left(a_2+d_0
\left(a_3+a_4 d_0\right)\right) \,. \ee Note further that \be P|_{a_4=0}
\sim a_3 (a_2 + a_3 d_0)^2 \,, \ee where $P_{H}|_{h=a_4=0}= a_3 (a_2
+ a_3 d_0)$ and $P_{M}|_{h= a_4 =0} =  (a_2 + a_3 d_0)$.

Now consider the Yukawa couplings, which arise from rank two enhanced points.  One obtains $\mathbf{10}\times\mathbf{\overline{5}}\times\mathbf{\overline{5}}$ couplings from $SO(12)$ points where $b_3=b_5=0$.  In our new variables, this corresponds to
\begin{equation}
SO(12): \qquad a_4=a_2+a_3d_0=0 \,.
\end{equation}
In particular, at these points both $P_H$ and $P_M$ vanish, and thus the matter fields on both participate in the Yukawa coupling.  This correctly reproduces the $\mathbf{10}_M\times \mathbf{\overline{5}}_M \times \mathbf{\overline{5}}_H$ coupling.

On the other hand, one obtains $\mathbf{10}\times\mathbf{10}\times \mathbf{5}$ couplings from $E_6$ points where $b_4=b_5=0$.  In our new variables, this is equivalent to
\begin{equation}
E_6: \qquad a_4=a_3=0 \,,
\end{equation}
and thus only the ${\bf 10}$ matter curve and the component $P_H$ of the ${\bf 5}$ matter curve participate.  This gives us precisely the $\mathbf{10}_M\times\mathbf{10}_M\times\mathbf{5}_H$ Yukawa coupling.



\subsection{Flux Quantization}
In Section 2.3.3 we discussed the quantization condition for bundles on a generic spectral surface ${\cal{C}_{\mathbf{10}}}$. Here we consider the split
${\cal{C}_{\mathbf{10}}}={\cal{C}}_{\mathbf{10}}^{(4)}+{\cal{C}}_{\mathbf{10}}^{(1)}$ so that
we have bundles $L_4$ and $L_1$ on ${\cal{C}}_{\mathbf{10}}^{(4)}$ and ${\cal{C}}_{\mathbf{10}}^{(1)}$ along with two ramification divisors, $r_4$ and $r_1$.  Furthermore, there are two projection maps
\begin{equation}
p_{\mathcal{C}_{\mathbf{10}}^{(1)}}:{\cal{C}}_{\mathbf{10}}^{(1)}\rightarrow S \,,\qquad p_{\mathcal{C}_{\mathbf{10}}^{(4)}}:{\cal{C}}_{\mathbf{10}}^{(4)}\rightarrow S\,.
\end{equation}

The condition that we need to impose there is that
\begin{equation}0=c_1(p_{\mathcal{C}_{\mathbf{10}}^{(4)}\,\ast}L_4)+c_1(p_{\mathcal{C}_{\mathbf{10}}^{(1)}\,\ast}L_1)=\left[p_{\mathcal{C}_{\mathbf{10}}^{(4)}\,\ast}c_1(L_4)-\frac{1}{2}p_{\mathcal{C}_{\mathbf{10}}^{(4)}\,\ast}r_4\right] + \left[p_{\mathcal{C}_{\mathbf{10}}^{(1)}\,\ast}c_1(L_1)-\frac{1}{2}p_{\mathcal{C}_{\mathbf{10}}^{(1)}\,\ast}r_1\right]\,.\end{equation}
We decompose $c_1(L_4)$ and $c_1(L_1)$ similarly as
\begin{equation}c_1(L_4)=\frac{1}{2}r_4+\gamma_4\,,\qquad c_1(L_1)=\frac{1}{2}r_1+\gamma_1\,,\end{equation}
and require that
\begin{equation}p_{\mathcal{C}_{\mathbf{10}}^{(4)}\,\ast}\gamma_4+p_{\mathcal{C}_{\mathbf{10}}^{(1)}\,\ast}\gamma_1=0\,.\end{equation}
To study the implications of this, let us compute $r_1$ and $r_4$.  Recalling that $X=\mathbb{P}({\cal{O}}\oplus K_S)$ is the ambient space and that $c_1(T_X)=2\sigma_{\infty}$ we have that
\begin{equation}\begin{split}c_1(T_{{\cal{C}}_{\mathbf{10}}^{(1)}}) &= \left[c_1(T_X)-[\mathcal{C}_{\mathbf{10}}^{(1)}]\right]|_{\mathcal{C}_{\mathbf{10}}^{(1)}} = \sigma_{\infty}|_{{\cal{C}}_{\mathbf{10}}^{(1)}} \\
c_1(T_{\mathcal{C}_{\mathbf{10}}^{(4)}}) &= \left[c_1(T_X)-[\mathcal{C}_{\mathbf{10}}^{(4)}]\right]|_{\mathcal{C}_{\mathbf{10}}^{(4)}}=\left(-2\sigma_{\infty}-\pi^{\ast}(\eta-5c_1)\right)|_{{\cal{C}}_{\mathbf{10}}^{(4)}}\,.
\end{split}\end{equation}
This means that
\begin{equation}\begin{split}
r_1 &= -\sigma\cap {\cal{C}}_{\mathbf{10}}^{(1)}=0 \\
r_4 &= \left(2\sigma+\pi^{\ast}(\eta-2c_1)\right)\cap {\cal{C}}_{\mathbf{10}}^{(4)}\,.
\end{split}\end{equation}
That $r_1$ is trivial is a natural consequence of the fact that it is a "1-sheeted" cover of $S$ and hence is unramified.  On the other hand, $r_4$ is even, a result that follows naturally from the fact that it is an even-sheeted cover of $S$.  This means that both $\gamma_1$ and $\gamma_4$ are integer quantized.

An important consequence of this integral quantization is, that we do not need to switch on the universal flux. This is in contrast to e.g. the case of the 5-sheeted unfactored cover, where the flux was half-integrally quantized and a nonzero universal flux had to be switched on.  In Appendix \ref{univii} we compute the chiral spectrum induced by the universal flux. In the next section we turn on non-universal fluxes
but turn off universal flux, which is allowed for the split ${\cal{C}_{\mathbf{10}}}={\cal{C}}_{\mathbf{10}}^{(4)}+{\cal{C}}_{\mathbf{10}}^{(1)}$.

\subsection{Nonuniversal Fluxes}

Let us now describe several ways to construct nonuniversal fluxes.  In each case, we start with a curve $\alpha_0$ in $S_{\rm GUT}$ that can be lifted to various sheets of ${\cal{C}}_{\mathbf{10}}$.  One trivial example of such a lift is obtained by pulling back $\alpha_0$ to $\mathcal{C}_{\mathbf{10}^{(1)}}$ via the projection $p_{\mathcal{C}_{\mathbf{10}^{(1)}}}$.  We call this curve $\tilde{\alpha}$,
\begin{equation}\tilde{\alpha} = p_{ \mathcal{C}_{\mathbf{10}}^{(1)}}^{\ast}\alpha_0\end{equation}
Using this we can construct a traceless flux as
\begin{equation}4\tilde{\alpha}-p_{\mathcal{C}_{\mathbf{10}}^{(4)}}^{\ast}p_{\mathcal{C}_{\mathbf{10}}^{(1)}\ast}\tilde{\alpha}\end{equation}

To do anything else, we need a curve $\alpha\in H_2(\mathcal{C}_{\mathbf{10}}^{(4)},\mathbb{Z})$ such that $p_{\mathcal{C}_{\mathbf{10}}^{(4)}\ast}\alpha = \alpha_0$.  In general, if $\alpha_0$ is defined by
\begin{equation}\alpha_0:\qquad F_{\alpha_0}=0,\qquad z_{S_{\rm GUT}}=0,\qquad U=0\end{equation}
for some defining equation $F_{\alpha_0}$ in $S_{\rm GUT}$ then we can construct a lift, $\alpha$, to a single sheet of ${\cal{C}}_{\mathbf{10}}^{(4)}$ as
\begin{equation}\alpha:  \qquad F_{\alpha_0}=0\qquad z_{S_{\rm GUT}}=0\qquad fV+gU=0\label{AlphaDef}\end{equation}
for some suitable $f$ and $g$ which must be tuned to ensure $\alpha\subset {\cal{C}}_{\mathbf{10}}^{(4)}$.  Using such an $\alpha$, we can construct two types of
traceless fluxes
\begin{equation} \label{AlphaTraceLess}
\ba
1. &\qquad 4\alpha - p_{\mathcal{C}_{\mathbf{10}}^{(4)}}^{\ast}p_{\mathcal{C}_{\mathbf{10}}^{(4)}\,\ast}\alpha \cr
2. &\qquad \alpha - p_{\mathcal{C}_{\mathbf{10}}^{(1)}}^{\ast}p_{\mathcal{C}_{\mathbf{10}}^{(4)}\,\ast}\alpha = \alpha-\tilde{\alpha} \,.
\ea
\end{equation}
The only other possibility is if we have multiple $\alpha_i$ in ${\cal{C}}_{\mathbf{10}}^{(4)}$ which satisfy $p_{\mathcal{C}_{\mathbf{10}}^{(4)}\,\ast}\alpha_i=\alpha_0$.  In that case, we can construct any sum
\begin{equation}\tilde{a}\tilde{\alpha} + \sum_i a_i \alpha_i\end{equation}
provided
\begin{equation}\tilde{a}+\sum_i a_i = 0 \,.\end{equation}
The net chirality induced on the ${\bf 10}$ matter curves equals the net  chirality induced  on the $ \overline{\bf 5}$ matter curves, and is given by
\be
N_{\bf 10}^{\rm total} = N_{\bf 5}^{\rm total} = \gamma \cdot \Sigma_{10}^{\rm total} \,.
\ee



\section{Fluxes and models with three generations}
\label{sec:compact}

In this section we give examples for fluxes that give rise to models with three generations and the correct Yukawa couplings.

\subsection{Three Generation Example}

In our first example  we consider switching on flux only in the ${\cal{C}}_{\mathbf{10}}^{(4)}$ part of the cover.
We need to ensure that no chiral generations are introduced on $\Sigma_{5,H}$.
This can be achieved by starting with an $\alpha_0\in H_2(S_{GUT},\mathbb{Z})$ that has the following intersections inside $S_{\rm GUT}$
\begin{equation}
\alpha_0\cdot_S \Sigma_{10}=m\ne 0 \,,\qquad \alpha_0\cdot_S P_{5,H}=0 \,.
\end{equation}
From this, we construct a curve $\alpha\in {\cal{C}}_{\mathbf{10}}^{(4)}$ as in (\ref{AlphaDef}) such that $\alpha$ covers $\alpha_0$ but does not intersect $U=0$.  Then, we build the traceless flux
\begin{equation}
\gamma = 4\alpha - p_{{\cal{C}}_{\mathbf{10}}^{(4)}\,\ast}p_{{\cal{C}}_{\mathbf{10}}^{(4)}}^{\ast}\alpha \,.
\end{equation}
If we evaluate $\gamma\cdot_{{\cal{C}}_{\mathbf{10}}^{(4)}}\Sigma_{10}$ the result will be $-m$.  Further, if we evaluate $\gamma\cdot_{{\cal{C}}_{\mathbf{10}}^{(4)}}\Sigma_{5,H}$ we are guaranteed to get 0 because $\alpha_0$ misses $P_{5,H}=0$.  The net  ${\bf 10}$ and net $\overline{\bf 5}$ chiralities have to agree, and thus this guarantees that $\gamma\cdot_{{\cal{C}}_{\mathbf{10}}^{(4)}}\Sigma_{5,M}=-m$.

\subsubsection{Sanity Check}

A necessary condition for this to work seems to be that $\alpha_0$ intersect $P_{5,M}$ in $S_{GUT}$. We can see that this has to be the case as follows: Recall that inside $S_{GUT}$, the classes of $P_{5,M}$, $P_{5,H}$, and $\Sigma_{10}$ are
\begin{equation}\begin{split}
[\Sigma_{10}]|_{S_{GUT}}&=\eta-5c_1 \\
[P_H]|_{S_{GUT}} &= 2\eta-7c_1 \\
[P_M]|_{S_{GUT}} &= \eta-3c_1 \,.
\end{split}\end{equation}
Now, suppose that $\alpha_0$ had vanishing intersection with both $P_H$ and $P_M$.  This would imply that
\begin{equation}\alpha_0\cdot_{S_{GUT}} [P_H-2P_M]=0\implies \alpha_0\cdot_{S_{GUT}} c_1=0 \,.
\end{equation}
This combined with requiring $\alpha_0\cdot_{S_{GUT}} P_H=0$ individually would force $\alpha_0\cdot_{S_{GUT}}\eta=0$ and hence force $\alpha_0$ to have vanishing intersection with all of $\Sigma_{10}$, $P_H$, and $P_M$ inside $S_{GUT}$.  So if $\alpha_0$ intersects $\Sigma_{10}$ but not $P_H$ inside $S_{GUT}$ it is forced to intersect $P_M$ at least once.


\subsubsection{Getting Three Generations}

Now let us turn our attention to the possible choices we have for $m$.  The condition that $\alpha_0$ misses $P_H$ requires
\begin{equation}
2\alpha_0\cdot_{S_{GUT}}\eta = 7\alpha_0\cdot_{S_{GUT}}c_1 \,,
\end{equation}
which further implies that
\begin{equation}
\alpha_0\cdot_{S_{GUT}}(\eta-5c_1) = -\frac{3}{2}\alpha_0\cdot_{S_{GUT}}c_1 \,.
\end{equation}
Because $\alpha_0$ is an integer class, this means that any such $\alpha_0$ has
\begin{equation}\alpha_0\cdot_{S_{GUT}}c_1=2n\,,\end{equation}
for some integer $n$.  From this it follows that
\begin{equation}\alpha_0\cdot_{S_{GUT}}\eta = 7n\end{equation}
and hence that
\begin{equation}\alpha_0\cdot_{S_{GUT}}\Sigma_{10} = -3n\,,
\qquad \alpha_0\cdot_{S_{GUT}}P_{5,M}=n \,.
\end{equation}

Now, suppose we construct $\alpha$ which covers $\alpha_0$ once but misses $U=0$.  In this case, $\gamma\cdot_{{\cal{C}}_{\mathbf{10}}^{(4)}}\Sigma_{10}=3n$, while $\gamma\cdot_{{\cal{C}}_{\mathbf{10}}^{(4)}}P_{5,M} = 4n-n=3n$.  Finally, we choose $n=1$ and this completes the task of finding the required spectrum. Note, that due to the structure of the factorized spectral cover, the presence of the correct Yukawa couplings is ensured automatically.


\subsubsection{Realization in Compact Geometry?}

Let us now turn to the explicit compact geometry constructed in \cite{Marsano:2009ym}, where $S_{\rm GUT} = dP_2$.
We review the topology of the base three-fold in Appendix D.
In this compact model the classes of the total matter curves are
\begin{equation}\begin{split}\Sigma_{10}&=2h-(e_1+e_2) \\
\Sigma_{5,H} &= 13h-5(e_1+e_2) \\
\Sigma_{5,M} &= 8h-3(e_1+e_2) \,.
\end{split}\end{equation}
Furthermore
\be
[F_Y] = e_1 - e_2 \,.
\ee
A curve $\alpha_0$ which satisfies $\alpha_0\cdot_{S_{GUT}}P_{5,H}=0$ and $\alpha_0\cdot_{S_{GUT}}\Sigma_{10}=-3$ is of the form
\begin{equation}\alpha_0 = 5h -be_1 - (13-b)e_2\,.\end{equation}
This cannot be symmetric in $e_1\leftrightarrow e_2$. We require that $\alpha_0$ is globally well-defined, i.e. arises from the intersection of a divisor in $B_3$ with $S_{\rm GUT}$. However, divisors in the compact model of \cite{Marsano:2009ym}, intersect  $S_{\rm GUT}$ in curves that are symmetric in $e_1$ and $e_2$. 
Hence if we want to use the $B_3$ constructed in \cite{Marsano:2009ym} as the base for our elliptically fibered Calabi-Yau four-fold then any $\alpha_0$ that we use when defining fluxes must be symmetric  in $e_1$ and $e_2$. So, in particular, this flux has unfortunately no realization in the  geometry of \cite{Marsano:2009ym}.

\subsection{Example with realization in compact setup}

We will now construct fluxes, which have a realization in the compact geometry of \cite{Marsano:2009ym}, meaning, in particular, that the class $\alpha_0$ inside $S_{\rm GUT}$ is symmetric in $e_1$ and $e_2$.
As in the first example, we will avoid matter on $\Sigma_{5,H}$ by building the fluxes from $\alpha_0$ satisfying $\alpha_0\cdot \Sigma_{5,H}=0$. These two requirements yield
\begin{equation}\alpha_0 = n\left[10h-13(e_1+e_2)\right] \,,
\end{equation}
which satisfies
\begin{equation}\alpha_0\cdot \Sigma_{10} = -6n \,,\qquad \alpha_0\cdot \Sigma_{5,M}=-8n\,.
\end{equation}
Even numbers like this are not promising for getting an odd number of generations.


Our general approach will be to look for traceless fluxes that can give three generations of $\mathbf{10}$'s and $\mathbf{\overline{5}}_M$'s.  In such a situation, it will be guaranteed that there is no net flux on the Higgs matter curves.
Our problem is that all of the relevant homological intersections inside $S_{\rm GUT}$ tend to be even.  What we need, then, are fluxes whose intersections in ${\cal{C}}$ do not  reduce to simple homological intersections.  For instance, consider the curve $\alpha$ in ${\cal{C}}_{\mathbf{10}}^{(4)}$ defined by (\ref{AlphaDef})
for some curve $\alpha_0$ in $S_{GUT}$.  Note that we must \emph{assume} that ${\cal{C}}_{\mathbf{10}}^{(4)}$ has been tuned so that this $\alpha$ is really inside ${\cal{C}}_{\mathbf{10}}^{(4)}$.

 With this assumption, the intersection of this curve with $\Sigma_{10}$ inside ${\cal{C}}_{\mathbf{10}}^{(4)}$ is given by the number of simultaneous solutions to
\begin{equation}F_{\alpha_0}=a_4=f=0\,,
\end{equation}
inside $S_{GUT}$.  Generically, this number will be zero because the number of points in the set $\alpha_0\cdot_{S_{GUT}} \Sigma_{10}$ which also satisfy $f=0$ vanishes.  If we tune $f$ appropriately, though, this number can be nonzero and less than the homological intersection $\alpha_0\cdot_{S_{GUT}}\Sigma_{10}$.  Indeed, we know of many instances in which expressions like $\alpha_0=f=0$ yield not a finite set of points but rather a curve which can intersect $\Sigma_{10}$.  The canonical example in the geometry of \cite{Marsano:2009ym}
 of this is $\alpha_0\sim W_3$ and $f\sim W_2$ in which case $\alpha_0=f=0$ defines a $\mathbb{P}^1$ in the class $h-e_1$.

Similarly, the intersection of $\alpha$ with the component $\Sigma_{5,M}^{(4)}$ of the $\mathbf{5}_M$ matter curve inside ${\cal{C}}_{\mathbf{10}}^{(4)}$ is given by the number of simultaneous solutions to
\begin{equation} F_{\alpha_0}=P_M=fd_0+g=0\,.\end{equation}
In general, this will also be less than the homological intersection $\alpha_0\cdot_{S_{GUT}}\Sigma_{5,M}$.

To make a traceless flux from this, we can take the construction 2. in (\ref{AlphaTraceLess})
\begin{equation}\hat{\alpha} = \alpha -p_{\mathcal{C}_{\mathbf{10}}^{(1)}}^{\ast}p_{\mathcal{C}_{\mathbf{10}}^{(4)}\,\ast}\alpha \,.
\end{equation}
The intersection of $p_{\mathcal{C}_{\mathbf{10}}^{(1)}}^{\ast}p_{\mathcal{C}_{\mathbf{10}}^{(4)}\,\ast}\alpha$ with $\Sigma_{5,M}^{(1)}$ will be just the homological intersection $\alpha_0\cdot_{S_{GUT}} \Sigma_{5,M}$.  In total, then
\begin{equation}\begin{split}\hat{\alpha}\cdot \Sigma_{10} &= \text{\# of points in }\alpha_0\cdot_{S_{GUT}}\Sigma_{10}\text{ with }f=0 \\
\hat{\alpha}\cdot(\Sigma_{5,M}^{(1)}+\Sigma_{5,M}^{(4)}) &= (-1)\times \text{\# of points in }\alpha_0\cdot_{S_{GUT}}\Sigma_{5,M}\text{ with }fd_0+g\ne 0 \,.
\end{split}\end{equation}
Further, $\hat{\alpha}\cdot \Sigma_{5,H}$ is just the difference of these two since the net number of $\mathbf{\overline{5}}$'s and $\mathbf{10}$'s must agree for traceless fluxes.

Unfortunately, the net chiralities on $\Sigma_{10}$ and $\Sigma_{5,M}$ appear with opposite sign here.  To fix this, we could look for another flux of the same general type.  A simpler option, however, is to consider a flux of the form
\begin{equation}\hat{\beta} = \left(4 p_{\mathcal{C}_{\mathbf{10}}^{(1)}}^{\ast}- p_{\mathcal{C}_{\mathbf{10}}^{(4)}}^{\ast}\right)\beta_0 \,,
\end{equation}
for some $\beta_0$ in $S_{GUT}$.  This will have intersections
\begin{equation}\begin{split}\hat{\beta}\cdot \Sigma_{10} &= -\beta_0\cdot_{S_{\rm GUT}}\Sigma_{10} \\
\hat{\beta}\cdot\Sigma_{5,M} &= 3\beta_0\cdot_{S_{\rm GUT}}\Sigma_{5,M}\,.
\end{split}\end{equation}

Again, let us realize this in the compact setup of \cite{Marsano:2009ym}.
Consider the choice for the flux curves
\begin{equation}\alpha_0=h \,,\qquad \beta_0=2(h-e_1-e_2) \,.
\end{equation}
Further, we suppose that $\alpha_0$ is reducible according to $h\rightarrow (h-e_1)+e_1$.  We expect the conditions $f=0$ and $fd_0+g=0$ to distinguish between these components.  In particular, we construct $f$ and $fd_0+g$ so that $f=\alpha_0=0$ contains a full curve in the class $e_1$ while $fd_0+g=\alpha_0=0$ contains a full curve in the class $h-e_1$.  In this case,
\begin{equation}\ba
\hat{\alpha}\cdot_{\cal{C}_{\mathbf{10}}} \Sigma_{10} &= 1 \cr
\hat{\alpha}\cdot_{\cal{C}_{\mathbf{10}}} \Sigma_{5,M} &= -3\cr
\hat{\beta} \cdot_{\cal{C}_{\mathbf{10}}} \Sigma_{10} &= 0 \cr
\hat{\beta}\cdot_{\cal{C}_{\mathbf{10}}} \Sigma_{5,M} &= 12 \,.
\ea
\ee
Now, define the total flux to be
\begin{equation}\hat{\gamma}=3\hat{\alpha} + \hat{\beta}\,,\end{equation}
which satisfies
\begin{equation}\hat{\gamma}\cdot_{\cal{C}}\Sigma_{10} = \hat{\gamma}\cdot_{\cal{C}}\Sigma_{5,M} = 3\,.
\end{equation}
By general reasoning, we are also guaranteed to have $\hat{\gamma}\cdot \Sigma_{5,H}=0$.

All we need, then, is to choose $f$ and $g$ appropriately and then tune ${\cal{C}}_{\mathbf{10}}^{(4)}$ so that it contains the curve $\alpha$.  One nice choice is the following.  Let us take
\begin{equation}\alpha_0 = W_1\,,\qquad f=W_3\,.
\end{equation}
Recall also that $d_0$ is in the class $c_1=3h-e_1-e_2$.  Since $f$ is in the class $h$ we see that $g+fd_0$, and hence also $g$ itself, is in the class $4h-e_1-e_2$.  Regardless of what $d_0$ turns out to be, we can use a judicious choice of $g$ to set
\begin{equation}g+fd_0 = W_4P_2(W_i)\,,\end{equation}
where $P_2(W_i)$ is a quadratic polynomial in $W_1,W_2,W_3$.  Now, $f=\alpha_0=0$ is not a collection of points but rather a curve in the class $e_1$.  Similarly, $g+fd_0=\alpha_0=0$ will include a curve in the class $h-e_1$.  We must choose $P_2(W_i)$ sufficiently generic that $P_2(W_i)=\alpha_0=0$ is a finite set of points so that there is no contribution to the intersection with $\Sigma_{10}$.


\subsection{Tuning ${\cal{C}}_{\mathbf{10}}^{(4)}$}

In our discussion so far, we have assumed that $\gamma \in \mathcal{C}_{\mathbf 10}^{(4)}.$  However, in order to ensure that this is the case, we need to suitably tune the coefficients of $\mathcal{C}_{\mathbf 10}^{(4)}.$
 If we want it to contain a curve of the form $\{fV+gU=0,F_{\alpha_0}=0\}$ then the cover should take the form
\begin{equation}\begin{split}-a_1d_0U^4+a_1U^3V+a_2U^2V^2+a_3UV^3+a_4V^4&=\left(fV+gU\right)\left(c_0 U^3+c_1U^2V+c_2UV^2+c_3V^3\right)\\
&\qquad+\alpha_0\left(\tilde{c}_0U^4+\tilde{c}_1U^3V+\tilde{c}_2U^2V^2+\tilde{c}_3UV^3+\tilde{c}_4V^4\right) \,,
\end{split}\end{equation}
which imposes the following constraints on the coefficients:
\begin{equation}\begin{split}
a_4 &= c_3f + \tilde{c}_4\alpha_0 \\
a_3 &= c_2f + c_3g+\tilde{c}_3\alpha_0 \\
a_2 &= c_1f + c_2g+\tilde{c}_2\alpha_0 \\
a_1 &= c_0 f + c_1 g + \tilde{c}_1\alpha_0 \\
a_0 &= -a_1d_0 = c_0g + \tilde{c}_0\alpha_0\,.
\end{split}\end{equation}
In particular, we are free to choose $c_i$, $\tilde{c}_j$, and $\alpha_0$ provided these choices satisfy the nontrivial constraint
\begin{equation}c_0 g + \tilde{c}_0\alpha_0 + d_0\left[c_0f + c_1g + \tilde{c}_1\alpha_0\right]=0\end{equation}
Said differently, we require $c_0g + \tilde{c}_0\alpha_0$ to admit $d_0$ as a factor.  In our explicit example above, it means that
\begin{equation}d_0\text{ divides }\left[c_0 \left(W_4P_2(W_i) - W_3d_0\right)+\tilde{c}_0 W_1\right]\end{equation}
or, in other words, that
\begin{equation}d_0\text{ divides }\left[c_0W_4P_2(W_i) + \tilde{c}_0W_1\right] \,.\end{equation}
For specific $d_0$ this is not hard to arrange.


\section{Towards vanishing of the tree-level $\mu-$term}
\label{sec:mu}

In section \ref{sec:constraints} we found that vanishing of the tree-level $\mu-$term
imposes a constraint on  monodromy group $G$ of the fundamental
spectral cover: \be G\in \{\mathbb{Z}_4,D_4,V\}\,.
\ee

In this section we first formulate a condition on coefficients of
the spectral cover to achieve such $G.$ Then we discuss the
technical difficulty which arises in trying to satisfy this
condition.

\subsection{Condition to get $G\in \{\mathbb{Z}_4,D_4,V\}$}
 Let us forget for a moment that the coefficients
of ${\cal{C}}_{\bf 10}^{(4)}$ are nontrivial sections and just
think about the equation for ${\cal{C}}_{\bf 10}^{(4)}$ as a
generic quartic
\begin{equation}f(s) := a_0 s^4 + a_1 s^3 + a_2s^2 + a_3s + a_4=0\end{equation}
for some elements $a_i$ of a particular function field $F$. We
suppose that the roots of $f(s)$ lie outside of $F$. The minimal
field extension of $F$ that contains all of the roots is called the
splitting field of $f$ and we shall denote it by $K$. Symmetries
that act nontrivially on $K$ but fix $F$ comprise the elements of
the Galois group of $f$. When the $a_i$ are taken to vary over
$S_{GUT}$, the monodromy group that is realized is generically
equivalent to this Galois group though it could in principle be a
subgroup.  In the end, we can always construct the antisymmetric
spectral surface and deduce from that precisely what monodromies are
realized.

Let us proceed then with a study of quartics and the symmetry
structure of their roots following the discussion of this topic in \cite{artin}. We assume that the quartic does not factor
since, as we argued in section 3, this is favored by
phenomenological constraints. This means that the Galois group acts
transitively. The transitive subgroups of $S_4$ are
\begin{equation}S_4, \ A_4,\  D_4,  \mathbb{Z}_4,  V\,,\end{equation}
where $A_4$ is the alternating group (subgroup of even permutations inside $S_4$), $D_4$ is the dihedral group, $\mathbb{Z}_4$ the cyclic group, and $V$ the Klein four group generated by even permutations of order 2
\begin{equation}V = \{1, (12)(34), (13)(24), (14)(23)\}\,.
\end{equation}
To discriminate among these, we need to study functions of the roots that are not quite symmetric.  Let us denote the roots by $u_1,u_2,u_3,u_4$.  The first example of such a function is the object $\delta$ defined by
\begin{equation}\delta = \prod_{i<j}(u_i-u_j)\,.
\end{equation}
Note that $\delta$ is simply the square root of the discriminant
\begin{equation}\delta^2 = a_4\,.
\end{equation}
If the discriminant is a square then $\delta$ can be written in $F$ and the Galois group $G$ can only contain those permutations that leave $\delta$ invariant.  In other words $a_4$ is a square if and only if $G\subset A_4$.
Of the transitive subgroups listed above, only $V$ is contained inside $A_4$.  This means that
\begin{equation}a_4\text{ is a square if and only if }G=V\text{ or }A_4\,.
\end{equation}

To discriminate among the others, let us define the objects
\begin{equation}\begin{split}
\beta_1 &= u_1u_3+u_2u_4 \\
\beta_2 &= u_1u_2+u_3u_4 \\
\beta_3 &= u_1u_4+u_2u_3\,.
\end{split}\end{equation}
The cubic polynomial
\begin{equation}g=(x-\beta_1)(x-\beta_2)(x-\beta_3)\,.
\end{equation}
is invariant under $S_4$ so can be written in terms of the
elementary symmetric polynomials $a_m$.  Further, the stabilizer of
any $\beta_i$ is easily seen to be of order 8.  For $\beta_1$, for
instance, the stabilizer is the subgroup
$\{1,(13),(24),(13)(24),(14)(23),(12)(34),(1234),(1432)\}$.  This is
one of the three conjugate $D_4$ subgroups of $S_4$.  As such, we
see that if $g$ factorizes in $F$, $G\subset D_4$.  Note further
that both $\mathbb{Z}_4$ and $V$ are contained inside $D_4$.  Summarizing all
of this, we can discriminate between various transitive subgroups of
$S_4$ using the following table
\begin{equation}\begin{array}{c|c|c} & a_4\text{ is a square} & a_4\text{ is not a square} \\ \hline
g\text{ is reducible} & G=V & G=D_4\text{ or }\mathbb{Z}_4 \\
g\text{ is irreducible} & G=A_4 & G=S_4\end{array}\end{equation}

Of course, we still have to write $g$ in terms of the $a_m$.  This is easily accomplished
\begin{equation}\label{cubic}g=a_0^3x^3-a_0^2a_2x^2+(a_0a_1a_3-4a_0^2a_4)x+
4a_0a_2a_4-a_1^2a_4-a_0a_3^2\end{equation}

This does not yet allow us to distinguish $D_4$ from $\mathbb{Z}_4$.  The set
$\{\lambda_i+\lambda_j\}$ splits into two orbits under the $D_4$
subgroup generated by $(1234)$ and $(13)$, namely
$\{\lambda_1+\lambda_2,\lambda_2+\lambda_3,\lambda_3+\lambda_4,\lambda_4+\lambda_1\}$
and $\{\lambda_1+\lambda_3,\lambda_2+\lambda_4\}$.  Note that this
is the same for $\mathbb{Z}_4$ so for our purposes it will not be so
important to distinguish them.

We conclude that the cubic polynomial (\ref{cubic}) must be
reducible to ensure the absence of a tree-level $\mu-$term.

\subsection{Difficulty with factoring $g$}
Let us try to factor $g$ defined in (\ref{cubic}). A simple way to
proceed is to set
\begin{equation}4a_0a_2a_4-a_1^2a_4-a_0a_3^2=0\,.
\end{equation}

We should remember that we impose (\ref{b1zero}), to ensure $b_1=0$.
Plugging this into $g$ yields
\begin{equation}g=a_1\left[a_1^2d_0^3 x^3 + a_1a_2 d_0^2 x^2 + \left(a_1a_3d_0 + 4a_1a_4d_0^2\right)x +\left[ a_1a_4 + d_0\left(4a_2a_4-a_3^2\right)\right]\right]\,.\end{equation}
Let us set
\begin{equation}a_1a_4+d_0(4a_2a_4-a_3^2)=0\,.\end{equation}
If we do this, then $g$ has $x$ as a factor and the monodromy group is
either $D_4$ or $\mathbb{Z}_4$ or $V$ as desired.

One potential problem is that we do not want $a_3$ and $a_4$ to have
a common factor because this would give an entire locus of $E_6$
singular fibers.  Instead, however, we can set
\begin{equation}d_0 = \delta a_4\,,\end{equation}
for
\begin{equation}\delta \sim 6c_1-\eta \sim t\,.\end{equation}
We then set
\begin{equation}a_1 = \delta\left(a_3^2-4a_2a_4\right)\,.\end{equation}

We can do this and plug into the equation for the antisymmetric
spectral surface.  We get
\begin{equation}\begin{split}{\cal{C}}_{\bar 5} &= b_0^3\left[s^{10}+3s^8c_2-s^7c_3+s^6(3c_2^2-3c_4)+s^5(-2c_2c_3+11c_5)\right.\\
&\quad + s^4(c_2^3-c_3^2-2c_2c_4)+s^3(-c_2^2c_3+4c_3c_4+4c_2c_5) \\
&\left.\quad + s^2(-c_2c_3^2+c_2^2c_4-4c_4^2+7c_3c_5)+s(c_3^3+c_2^2c_5-4c_4c_5)+(c_2c_3c_5-c_3^2c_4-c_5^2)\right] \\
&=\left(a_3^2s\delta(1+a_4s\delta)-a_2(1+2a_4s\delta)^2\right) \\
& \quad \times \left[a_3+4a_4^2\delta+a_3^2s^2\delta(1+a_4s\delta)^2-a_2s(1+a_4s\delta)(1+2a_4s\delta)^2\right] \\
&\quad \times \left[a_2(1-2a_4s\delta)^2(-1+a_4s\delta)^2+\delta(a_4^3\delta+a_3a_4(1-a_4s\delta)-a_3^2s(-1+a_4s\delta)^3\right] \,,
\end{split}\end{equation}
where $c_i=b_i/b_0$ and we substituted in for $a_m$, etc.  The third
term in the product here is the one that we had before from the
factorization ${\cal{C}}_5\rightarrow {\cal{C}}_{\mathbf{10}}^{(4)}+{\cal{C}}_{\mathbf{10}}^{(1)}$.  The
first two are the newly factored sixth order polynomial that
comprises the antisymmetric spectral surface associated to
${\cal{C}}_{\mathbf 10}^{(4)}$.

We should also see the $P_H$ polynomial factor (and the $P_M$
polynomial not factor).  Indeed, we find that
\begin{equation}
P_H\rightarrow a_2(a_3+4a_4^2\delta)\,,\qquad P_M\rightarrow a_2+a_4\delta (a_3+a_4^2\delta)\,.
\end{equation}
Note however that the classes of the two components into which $P_H$
splits are completely fixed and are linear combinations of $c_1$ and
$t,$ which satisfy $F_Y \cdot c_1= F_Y \cdot t =0$. Therefore, one
cannot get non-zero restriction of $[F_Y]$ to either of these
two components of $P_H.$

We can try to remedy the situation by requiring 
 $g$ to have a root at a generic $x=x_0$.  This amounts to
solving
\begin{equation}a_1^2d_0^3x_0^3+a_1a_2d_0^2x_0^2+(a_1a_3d_0+4a_1a_4d_0^2)x_0 + (a_1a_4+d_0(4a_2a_4-a_3^2))=0 \,.
\end{equation}
If we notice from (\ref{intermsofa}) that
\begin{equation}a_1a_4 = a_3(a_2+a_3d_0)-P_H \,,\end{equation}
and plug this into the order $x_0$ term we find
\begin{equation}P_H = \left(a_2+a_1d_0x_0\right)\left(a_3+d_0(4a_4+a_1d_0x_0^2)\right) \,.
\end{equation}
Because $P_H$ depends on $a_1,a_2,a_3,a_4,d_0$, this is a nontrivial condition to solve.
However, we see that whenever it is solved the Higgs matter curve necessarily factorizes.
From here it again seems that the classes of the two components into
which $P_H$ splits are completely fixed.  However, we must be a bit
careful about that.  Note that $[x_0]=-2c_1$ so to work with
effective classes we should really talk about $y_0\sim 1/x_0$ with
$[y_0]=2c_1$.  In terms of $y_0$,  (\ref{intermsofa}) becomes
\begin{equation}\label{important}y_0^3P_H =
\left(a_2y_0 +
a_1d_0\right)\left[y_0^2\left(a_3+4d_0a_4\right)+a_1d_0^2\right]\,.
\end{equation}
Any consistent solution of $P_H=a_3(a_2+a_3d_0)-a_1a_4$ and
$g(x_0)=0$ will be such that $y_0^3$ divides the right hand side of
(\ref{important}). However, it is not completely determined how this
division takes place.  If $y_0$ is a product
\begin{equation}y_0\sim y_1 y_2 \,,
\end{equation}
we could adjust the classes of the factors of $P_H$ by choosing
things so that each of the factors above contains different powers
of $y_1$ and $y_2$. 
Doing so, without rendering our four-fold unacceptably singular, seems difficult thus far. 
We hope to return to this issue in the future.

\section*{Acknowledgements}

We thank  V.~Bouchard, J.~Heckman, C.~Vafa and M.~Wijnholt for valuable discussions.  We would also like to thank the organizers of the Strings 2009 conference in Rome for providing a wonderful venue in which to complete this work.  
JM is grateful to the string theory group at the University of Amsterdam for their hospitality as the second version of this paper was being completed.  The work of JM and SSN was supported by John A. McCone Postdoctoral Fellowships.
The work of NS was supported in part by the DOE-grant DE-FG03-92-ER40701.

\newpage

\appendix

\section{Computing the Spectrum with a Particular Representative for universal $\gamma$}
\label{universal}

To see what happens when $P$ becomes reducible into components let us pick a representative for universal
flux $\gamma$ and compute the spectrum of $\mathbf{\overline{5}}$'s.  We know that $\gamma_u$ is in the class
\begin{equation}\gamma_u = {{\cC}_{10}}\cap \left(5\sigma-\pi^{\ast}\eta+5\pi^{\ast}c_1\right)\,.
\end{equation}
A nice representative, $\tilde{\gamma}$, of $5\sigma-\pi^{\ast}\eta+5\pi^{\ast}c_1$ is simply
\begin{equation}\tilde{\gamma}=\frac{V^5}{b_0}\,.
\end{equation}

Let's start by using $\tilde{\gamma}$ to compute the net chirality on the $\mathbf{10}$ curve.  For this, we need to compute the intersection of $\tilde{\gamma}$ with $\Sigma_{10}$, which is the intersection of
\begin{equation}\begin{split}0 &= U \\
0 &= {\cal{F}}_{{\cC}_{10}}\end{split}\end{equation}
or equivalently the intersection of $U=0$ with $b_5=0$.  Because $U=0$, $V$ cannot be zero so we get only $(-1)$ times the intersection of $U=b_5=0$ with $b_0=0$.  This is simply the intersection of the $b_5=0$ and $b_0=0$ curves inside ${S_{\rm GUT}}$ so that we obtain
\begin{equation}n_{\mathbf{10}}-n_{\mathbf{\overline{10}}} = -\eta\cdot_{S_{\rm GUT}} (\eta-5c_1)\,.
\end{equation}

Now, let us turn to the net chirality on the $\mathbf{\overline{5}}$ curve.  We need to compute
\begin{equation}\tilde{\gamma}\cap \Sigma_5\,,\end{equation}
where $\Sigma_5$ is obtained from the intersection of the following two equations
\begin{equation}\begin{split}0 &= b_3U^2 + b_5V^2 \\
0 &= b_0 U^4 + b_2 U^2 V^2 + b_4 V^4\,.
\end{split}\end{equation}
In particular, we get 5 times the intersection of $\Sigma_5$ with $V=0$ minus 1 times the intersection of $\Sigma_5$ with $b_0=0$.  For the intersection of $\Sigma_5$ with $V=0$ we find $b_0=b_3=0$.  This contribution is then
\begin{equation}5\eta\cdot_{S_{\rm GUT}} \left(\eta-3c_1\right)\label{firstcont}\end{equation}
For the intersection of $\Sigma_5$ with $b_0=0$ we have already counted those points with $V=0$.  For those points without $V=0$ we can set $V=1$ to obtain the following equations for $\Sigma_5$
\begin{equation}\begin{split}0 &= b_3 u^2 + b_5 \\
0 &=b_0 u^4 + b_2 u^2 + b_4\end{split}\end{equation}
By solving the first equation for $u$ and plugging into the second, we see that this part of $\Sigma_5$ is a double cover of the curve in ${S_{\rm GUT}}$ defined by $P=0$ with
\begin{equation}P=b_0b_5^2-b_2b_3b_5+b_3^2b_4\end{equation}
which is singular at $b_3=b_5=0$ where it intersects itself.  We should perform a suitable resolution of $\Sigma_5$ in which the two sheets are separated and proceed from there.  Intersecting with $b_0=0$ then yields twice the intersection number of the curve $b_0=0$ with $P=0$ inside ${S_{\rm GUT}}$
\begin{equation}2\times \eta\cdot_{S_{\rm GUT}} \left(3\eta-10c_1\right)\end{equation}
Subtracting this from the first contribution \eqref{firstcont} we find the net result
\begin{equation}5\eta\cdot_{S_{\rm GUT}}\left(\eta-3c_1\right)-2\times \eta\cdot_S \left(3\eta-10c_1\right)=-\eta\cdot_{S_{\rm GUT}} \left(\eta -5c_1\right)\,.
\end{equation}

\section{Matter Curves for $4+1$ Factorization}
\label{app:Matter41}

In this appendix we provide the details for the analysis of matter curves in all components of $\mathcal{C} \cap \tau \mathcal{C}$ in the factorization $\mathcal{C} = {\cal{C}}_{\mathbf{10}}^{(1)} \cup {\cal{C}}_{\mathbf{10}}^{(4)}$.


\subsection{${\cal{C}}_{\mathbf{10}}^{(1)}\cap\tau{\cal{C}}_{\mathbf{10}}^{(1)}$}

We start with the component of ${\cal{C}}_{\mathbf{10}}^{(1)}$ that is invariant under $V\rightarrow -V$.  This is simply $V=d_0=0$ and is in the class
\begin{equation}{\cal{C}}_{\mathbf{10}}^{(1)}\cap [V] = \sigma_{\infty}\cdot\sigma_{\infty}=\sigma_{\infty}\cdot \pi^{\ast}c_1\end{equation}

In summary, we get contributions to various components of ${\cal{C}}\cap\tau{\cal{C}}$ from ${\cal{C}}_{\mathbf{10}}^{(1)}\cap \tau{\cal{C}}_{\mathbf{10}}^{(1)}$ as follows
\begin{equation}\begin{array}{
c|c|c}\text{Component} & \text{Class} & \text{Equations} \\ \hline
\mathbf{10}\text{ Matter Curve} & \cdot & \cdot \\ \hline
\mathbf{5}\text{ Matter Curve} & \cdot & \cdot \\ \hline
\text{Intersection at Infinity} & \sigma_{\infty}\cdot \pi^{\ast}c_1
& V=0, d_0=0 \\ \hline
\end{array}\end{equation}


\subsection{${\cal{C}}_{\mathbf{10}}^{(1)}\cap \tau{\cal{C}}_{\mathbf{10}}^{(4)}$}

We now turn to the intersection of ${\cal{C}}_{\mathbf{10}}^{(1)}$ with the $V\rightarrow -V$ image of ${\cal{C}}_{\mathbf{10}}^{(4)}$.  Plugging $V=-d_0U$ into $\tau{\cal{C}}_{\mathbf{10}}^{(4)}$ we get
\begin{equation}-U^4d_0^2(a_2+d_0a_3+d_0^2a_4)\equiv -U^4d_0^2 P_M\,.\end{equation}
We cannot get a solution from $U=0$ but we get a solution of multiplicity two from $V=d_0=0$ and a solution of multiplicity 1 from $V=-d_0U$ and $P_M=0$.  This corresponds to the decomposition
\begin{equation}{\cal{C}}_{\mathbf{10}}^{(1)}\cap \tau{\cal{C}}_{\mathbf{10}}^{(4)} = \left[\sigma_{\infty}\cdot \pi^{\ast}(\eta-3c_1)\right]+\left[2\sigma_{\infty}\cdot \pi^{\ast}c_1\right]\,.\end{equation}
The first of these is a $\mathbf{5}$ matter curve while the second is part of an "intersection at infinity", which we neglect.

We can rewrite the class of the $\mathbf{5}$ matter curve as
\begin{equation}\sigma\cdot \pi^{\ast}(\eta-3c_1) + \pi^{\ast}(c_1)\cdot \pi^{\ast}(\eta-3c_1)\end{equation}
where we understand the first term as coming from the intersection of $U=-V/d_0$ with $P_M=0$ and the second as coming from the intersection of $d_0=0$ with $P_M=0$.

In summary, we get contributions to various components of ${\cal{C}}\cap\tau{\cal{C}}$ from ${\cal{C}}_{\mathbf{10}}^{(1)}\cap \tau{\cal{C}}_{\mathbf{10}}^{(4)}$ as follows
\begin{equation}\begin{array}{c|c|c}\text{Component} & \text{Class} & \text{Equations} \\ \hline
\mathbf{10}\text{ Matter Curve} & \cdot & \cdot \\ \hline
\mathbf{5}\text{ Matter Curve} & \sigma \cdot\pi^{\ast}(\eta-3c_1)+\pi^{\ast}(c_1)\cdot\pi^{\ast}(\eta-3c_1) & V+d_0U=0, P_M=0 \\ \hline
\text{Intersection at Infinity} & 2\sigma_{\infty}\cdot \pi^{\ast}c_1 & V=0, d_0=0 \\ \hline
\end{array}\end{equation}

We get a similar contribution also from $\tau{\cal{C}}_{\mathbf{10}}^{(1)}\cap {\cal{C}}_{\mathbf{10}}^{(4)}$.


\subsection{${\cal{C}}_{\mathbf{10}}^{(4)}\cap \tau{\cal{C}}_{\mathbf{10}}^{(4)}$}

We finally turn to the components of ${\cal{C}}_{\mathbf{10}}^{(4)}$ that is invariant under $V\rightarrow -V$.  These satisfy
\begin{equation}UV\left(a_1U^2+a_3V^2\right)=0\,,\qquad -a_1d_0U^4+a_2U^2V^2+a_4V^4=0\,.\end{equation}

The first part of the solution is $U=a_4=0$.  This is in the class
\begin{equation}\sigma\cap {\cal{C}}_{\mathbf{10}}^{(4)} = \sigma\cap \pi^{\ast}(\eta-5c_1)\end{equation}

The next component is $V=a_1d_0=0$.  This is in the class
\begin{equation}V\cap {\cal{C}}_{\mathbf{10}}^{(4)}=V\cap \pi^{\ast}(\eta-c_1)=\sigma_{\infty}\cap \pi^{\ast}(\eta-c_1)\end{equation}

What remains are the intersection of ${\cal{C}}_{\mathbf{10}}^{(4)}$ with solutions to
\begin{equation}a_1U^2+a_3V^2\end{equation}
This has a solution with multiplicity two along $a_1=V=0$ which is in the class
\begin{equation}2\sigma_{\infty}\cdot\pi^{\ast}(\eta-2c_1)\end{equation}
What remains constitutes a $\mathbf{5}$ matter curve and is in the class
\begin{equation}\left[2\sigma + \pi^{\ast}(\eta-2c_1)\right]\cdot\left[4\sigma+\pi^{\ast}(\eta-c_1)\right] - 2\sigma_{\infty}\cdot\pi^{\ast}(\eta-2c_1) = 2\sigma\cdot \pi^{\ast}(2\eta-7c_1) + \pi^{\ast}(\eta-2c_1)\cdot\pi^{\ast}(\eta-3c_1)\end{equation}
Strictly speaking, the equations for this matter curve are
\begin{equation}a_1U^2+ a_3V^2=-a_1d_0U^4+a_2U^2V^2+a_4V^4=0\end{equation}
less the $a_1=V=0$ component.  We can understand how the homology class above arises as follows.  If we assume that $a_1\ne 0$ then we can solve for $U$ in the first equation to obtain
\begin{equation}U = \pm iV\sqrt{\frac{a_3}{a_1}}\label{UpmVsol}\end{equation}
Plugging back into ${\cal{C}}_{\mathbf{10}}^{(4)}$ then yields
\begin{equation}-(a_3^2d_0+a_3a_2-a_4a_1)\frac{V^4}{a_1}\equiv - P_H\frac{V^4}{a_1}=0\end{equation}
Since we assumed $a_1\ne 0$, from which it follows that $V\ne 0$, this component can be described by \eqref{UpmVsol} along with $P_H=0$ and is in the class
\begin{equation}2\sigma\cdot\pi^{\ast}(2\eta-7c_1)\end{equation}
We must also study the "points" at infinity where $a_1\rightarrow 0$.  In this neighborhood, we can instead solve the above equations as $a_1 = -a_3V^2/U^2$ and plug back into to ${\cal{C}}_{\mathbf{10}}^{(4)}$ to obtaon
\begin{equation}a_1=-\frac{a_3V^2}{U^2}\qquad V^2\left[(a_3d_0+a_2)U^2+a_4V^2\right]=0\end{equation}
The overall factor of $V^2$ in the second equation is capturing the $a_1=V=0$ "intersection at infinity" component that we already considered so we drop this.  Now, if $a_1\rightarrow 0$ due to $a_3$ becoming small, we see that we approach smooth points on the matter curve above that are away from $V=0$.  On the other hand, $a_1$ can become small due to $V$ becoming small if we move toward points in $S$ with $a_1=a_2+a_3d_0=0$.  These points are the ones that are accounted for homologically by the factor
\begin{equation}\pi^{\ast}(\eta-2c_1)\cdot\pi^{\ast}(\eta-3c_1)\end{equation}

In summary, we get contributions to various components of ${\cal{C}}\cap\tau{\cal{C}}$ from ${\cal{C}}_{\mathbf{10}}^{(4)}\cap \tau{\cal{C}}_{\mathbf{10}}^{(4)}$ as follows
\begin{equation}\begin{array}{c|c|c}\text{Component} & \text{Class} & \text{Equations} \\ \hline
\mathbf{10}\text{ Matter Curve} & \sigma\cdot\pi^{\ast}(\eta-5c_1) & U=0, a_4=0 \\ \hline
\mathbf{5}\text{ Matter Curve} & 2\sigma\cdot\pi^{\ast}(2\eta-7c_1) & U=\pm iV(a_3/a_1)^{1/2}, P_H=0 \\
& +\pi^{\ast}(\eta-2c_1)\cdot\pi^{\ast}(\eta-3c_1) & a_1=0, a_2+d_0a_3=0 \\ \hline
\text{Intersection at Infinity} & \sigma_{\infty}\cdot\pi^{\ast}(3\eta-5c_1) & V=0, a_1=0 \\ \hline
\end{array}\end{equation}

\section{Universal Fluxes for $4+1$ split}
\label{univii}

Here we find the chiral spectrum for the universal flux for the split ${\cal{C}_{\mathbf{10}}}={\cal{C}}_{\mathbf{10}}^{(4)}+{\cal{C}}_{\mathbf{10}}^{(1)}.$

There is the standard type of universal flux, as discussed in \cite{Donagi:2009ra, Marsano:2009ym}, which is present also in for unfactorized spectral covers
\begin{equation}\gamma_u = {\cal{C}}\cap (5\sigma_{\infty}-\pi^{\ast}\eta)
\end{equation}
This flux is generically present and by construction traceless
\begin{equation}
p_{C\,\ast} \gamma_u =0 \,.
\end{equation}


For a factorized spectral cover, $\gamma_u$ can be separated into a component inside ${\cal{C}}_{\mathbf{10}}^{(4)}$ and a component inside ${\cal{C}}_{\mathbf{10}}^{(1)}$
\begin{equation}
\gamma_u=\gamma_{u,1}+\gamma_{u,4} \,,
\end{equation}
which in turn are given by
\begin{equation}
\gamma_{u,1}={\cal{C}}_{\mathbf{10}}^{(1)}\cap(5\sigma_{\infty}-\pi^{\ast}\eta)\,,\qquad
\gamma_{u,4}={\cal{C}}_{\mathbf{10}}^{(4)}\cap (5\sigma_{\infty}-\pi^{\ast}\eta) \,.
\label{gammau1u4}
\end{equation}

Let's first integrate $\gamma_u$ over the $\mathbf{10}$ matter curve.  Because $\Sigma_{10}$ is contained entirely within ${\cal{C}}_{\mathbf{10}}^{(4)}$ we get
\begin{equation}({\cal{C}}_{\mathbf{10}}^{(4)}\cap\sigma)\cdot_{{\cal{C}}_{\mathbf{10}}^{(4)}} ({\cal{C}}_{\mathbf{10}}^{(4)}\cap 5\sigma_{\infty}-\pi^{\ast}\eta)=-\eta\cdot_{S_{GUT}} (\eta-5c_1)\end{equation}
which is the same result as always. Likewise, the {\bf 5} matter curve has the same induced chirality, as expected.

\section{Review of Three-fold base in the compact model}
In \cite{Marsano:2009ym} we constructed compact Fano three-folds $X$ and $\tilde X$ which can be used
as a base of elliptically fibered Calabi-Yau four-fold. Here we briefly review this construction
and summarize the topology of $X$ and $\tilde X.$

\subsection{Construction}
Let $Z=\mathbb{P}^3$ with homogenous coordinates $[Z_0,Z_1,Z_2,Z_3]$.  The canonical class is given in terms of the hyperplane class $H$ as
\be
K_Z = - 4 H \,.
\ee
Inside $\mathbb{P}^3$, we consider the nodal curve ${\cal{C}}$ defined by the equations
\be
\ba
Z_4Z_1Z_2+(Z_1+Z_2)^3&=0 \cr
            Z_3 &=0\,.
\ea
\ee
Alternatively, this can be written in affine coordinates $z_i$ as
\begin{equation}
{\cal{C}}=\left\{[z_1,z_2,0,1]\,|\, z_1z_2+(z_1+z_2)^3=0\right\}\cup\left\{[1,-1,0,0]\right\}\,.
\label{Ceqns}\end{equation}
In what follows, we will typically consider the affine patch $[z_1,z_2,z_3,1]$ of $\mathbb{P}^3$ since this contains all of ${\cal{C}}$ except for a single ``point at infinity".  As clear from \eqref{Ceqns}, ${\cal{C}}$ exhibits a singular point at $[0,0,0,1]$ which is of the form $z_1z_2=z_3=0$.

The first step in constructing our three-fold is to blow up along ${\cal{C}}$ to obtain the three-fold $Y$ with the blow-down map
\be
\psi: Y \rightarrow Z \,.
\ee
In coordinates this can be described by considering $\mathbb{C}^3\times\mathbb{P}^1$ in the $Z_4=1$ patch with homogeneous coordinates $[V_0,V_1]$ on the new $\mathbb{P}^1$, which we shall hereafter denote by $\mathbb{P}^1_V$.  The blow-up is then defined in this patch by the equation
\begin{equation}
Y\,:\qquad V_0\left(z_1z_2+(z_1+z_2)^3\right)=V_1z_3 \,.
\label{blowup1}
\end{equation}
From \eqref{blowup1}, we see that the resulting three-fold exhibits a singular point at $\{(z_1,z_2,z_3),[V_0,V_1]\}=\{(0,0,0),[1,0]\}$. Let us  pass to an
affine patch covering the north pole $v_0\not=0$ of $\mathbb{P}^1_V$. Then defining again $u = v_1 /v_0$ the equation \eqref{blowup1} in fact becomes
\begin{equation}
[z_1z_2+(z_1+z_2)^3]=uz_3\,,
\end{equation}
so that near the singular point it behaves like
\begin{equation}
z_1z_2=u  z_3\,.
\end{equation}
We recognize this as a conifold singularity.

The divisor classes in $Y$ are the exceptional divisor $Q$, which is a $\mathbb{P}^1$-bundle over ${\cal C},$
 and $\psi^* (H)  = Q +(H-Q)$. The canonical class is
\be
K_Y = \psi^* (K_Z) + Q = - 4 H + Q \,.
\ee

The final step is to blow-up the conifold singularity in $Y$ by
\be
\phi: X\rightarrow Y \,.
\ee
To do this, we move to a local patch covering the north pole of $\mathbb{P}^1_V$ with coordinates $(z_1,z_2,z_3,u=v_1/v_0).$  Let us blow up the origin of this $\mathbb{C}^4$ by gluing in a $\mathbb{P}^3_W$ with homogeneous coordinates $[W_1,W_2,W_3,W_4]$ and restrict to $z_1z_2=z_3u$ and its smooth continuation, $W_1W_2=W_3W_4$, at the origin.  In the end, the three-fold takes the following form in this local patch
\be\label{XDef}
\ba
X_1 = & \left\{(z_1,z_2,z_3,v_1;W_1,W_2,W_3,W_4)\in \mathbb{C}^4\times \mathbb{P}^3_W\,:\quad \right.\cr
& \left.  \qquad \qquad
(z_1,z_2,z_3, u) \in [W_1,W_2,W_3,W_4] \,, \quad
 z_1 z_2 = z_3  u \,,\quad W_1 W_2 = W_3 W_4
  \right\}\,.
  \ea
\ee

We can identify the two $\mathbb{P}^1$'s with the submanifolds
\be\label{P1s}
\mathbb{P}^1_{(1)}:\quad W_2=W_4=0\,,\qquad
\mathbb{P}^1_{(2)}:\quad W_2=W_3=0\,.
\ee
Note that in this local patch it is not possible to see that these $\mathbb{P}^1$s are in the same class in $X$.
It is however clear from the global topology of $X$ since their intersections with all divisors are equivalent.
The canonical class of $X$ is
\be
K_X = - 4 H + (D + E )+  E \,,
\ee
where the exceptional divisor is
\be
\phi^* Q = D + E  \,.
\ee

The curve $G$ is a $(-1,-1)$ curve because it is an exceptional $\mathbb{P}^1$ so that we can   flop it to obtain a new three-fold, $\tilde{X}$, depicted in figure \ref{fig:GlobalThree}.  The divisors $D$ and $E$ of $X$ carry over to new divisors $D'$ ad $E'$ in $\tilde{X}$.  The canonical class also follows simply from $K_X$ as
\be\label{KXtilde}
K_{\tilde{X}} = - 4 H + D' + 2 E' \,.
\ee
The resulting three-fold $\tilde{X}$ has  the desired property that the two curves $\ell-G'$ are distinct in $H_2(E',\mathbb{Z})$ but are nonetheless equivalent in $H_2(\tilde{X},\mathbb{Z})$ so that they satisfy the condition for existence of a suitable hypercharge flux.
\begin{figure}
\begin{center}
\epsfig{file=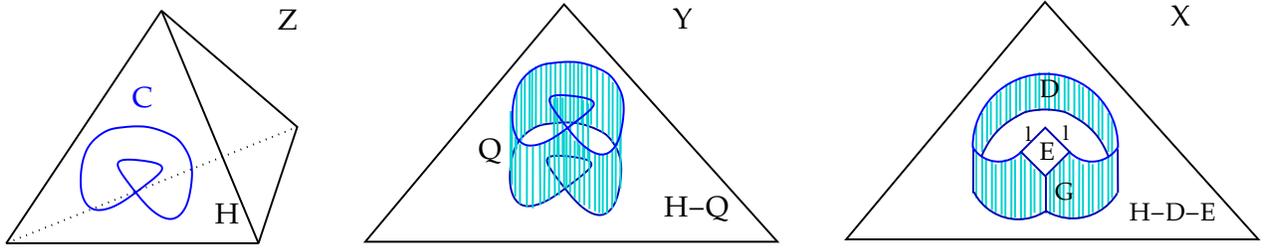,width=1\textwidth}
\caption{Global Construction of Threefold: blowups.}
\label{fig:Global12}
\end{center}
\end{figure}

\subsubsection{Topology of $X$}
Let us first summarize the topology of $X.$
As a basis of $H_2(X,\mathbb{Z})$, we take the curve $\ell_0$, which descends from the unique generator of $H_2(\mathbb{P}^3,\mathbb{Z})$, as well as the curves $\ell$ and $G$ depicted in figure \ref{fig:Global12}.
A useful basis of divisors is $H$, $E$ and $H-D-E$. Their topology is
\be
\ba
H &\cong dP_3\cr
E &\cong \mathbb{P}^1 \times \mathbb{P}^1 \cr
H-D-E &\cong \mathbb{P}^2 \,.
\ea
\ee
The intersection numbers with various divisors are given by the following table
\begin{center}
\begin{tabular}{|c||r|r|r|r|}
\hline
        &$H$    & $D$   &   $E$  \cr\hline\hline
$\ell_0$    & $+1$& $0$& $0$ \cr\hline
$\ell$      & $0$&$ +1$ & $-1$ \cr\hline
$G$     & $0$& $-2$ & $1$  \cr\hline
\end{tabular}
\end{center}

\begin{figure}
\begin{center}
\epsfig{file=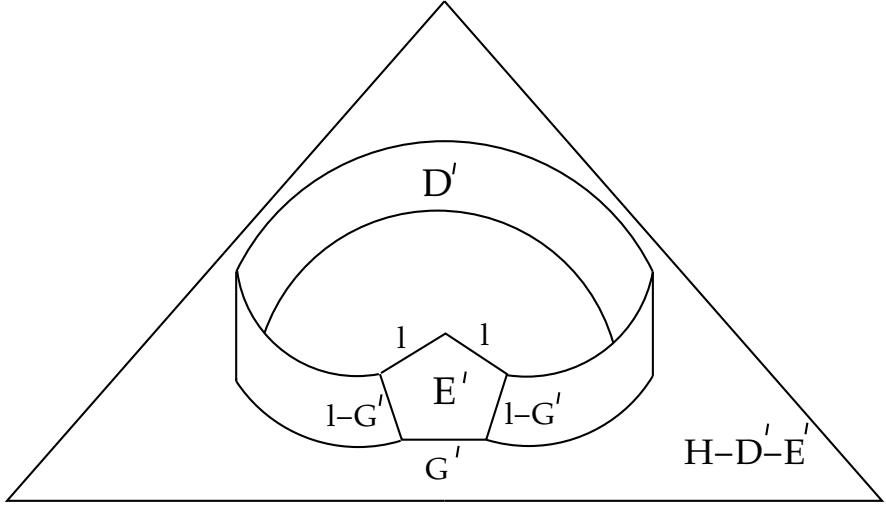,width=.7\textwidth}
\caption{Final  three-fold $\tilde{X}$}
\label{fig:GlobalThree}
\end{center}
\end{figure}


The intersections of divisors with one another is furthermore
\begin{equation}\label{DoubleIntBF}
\begin{array}{|c||c|c|c|}
\hline
& H & E & D \\ \hline\hline
H & \ell_0 & 0 & 3(\ell+G) \cr\hline
E & 0 & -2\ell & 2\ell \cr\hline
D & 3(\ell+G) & 2\ell & -3\ell_0 + 12(\ell+G) - 2\ell \cr\hline
H-D-E & \ell_0-3(\ell+G) & 0 & 3\left(\ell_0-3(\ell+G)\right)\cr\hline
\end{array}
\end{equation}
from which the following non-vanishing triple-intersections follow
\be
\ba
H^3 &= 1 \cr
D^3 &= -14 \cr
E^3 &= 2 \cr
D^2H &= -3 \cr
D^2E &= 2 \cr
E^2 D &= -2 \,.
\ea
\ee
Let us further recall the basis of holomorphic sections for $X$:

\be\label{SectionsX}
\begin{array}{|l||l| }\hline
\text{Holomorphic Section} & \hbox{Divisor class} \cr\hline\hline
Z_4 & H\cr\hline
Z_{1,2} & (H-E) + E =H \cr\hline
Z_3 & (H-D-E) + (D+E)= H \cr\hline
W_{1,2,3} & H-E\cr\hline
W_4 & 3H -D-2E \cr\hline
V_1 & ( 3H - D -2E) + E = 3 H - D - E \cr\hline
V_0 & H-D-E\cr\hline
\end{array}
\ee

Note that for $S_{\rm GUT}=X$ we find $t=-c_1({\cal N}S_{\rm GUT})=-E^2\vert_{E}=l_1+l_2$ where we use that
${\cal N}S_{\rm GUT}=E$ and the fact that the class $2l$ in $X$ restricts to the class
$l_1+l_2$ in $E=\mathbb{P}^1\times\mathbb{P}^1.$

\subsubsection{Topology of $\tilde X$}
Let us review the topology of $\tilde{X}$, including the topology of various divisors and the intersection tables for divisors and curves.  We start with a discussion of several interesting divisor classes.  The divisor $H$, which was a $dP_3$ before the flop, remains a $dP_3$ because it is unaffected by the flop.  From the viewpoint of $H-D-E=\mathbb{P}^2$, however, the flop corresponds to blowing up a point so that $H-D'-E'$ becomes a $dP_1$.  Similarly, from the viewpoint of $E=\mathbb{P}^1\times\mathbb{P}^1$,
the flop effectively blows up a point so that $E'$ is simply $dP_2$.
Finally the divisor $D'$ is the Hirzebruch surface $\mathbb{F}_4$.
\be\label{DivBasis}
\ba
H &\cong dP_3 \cr
E' &\cong dP_2\cr
D' &\cong\mathbb{F}_4 \cr
H-D'-E' &\cong dP_1 \,.
\ea
\ee

As a basis of $H_2(\tilde{X},\mathbb{Z})$, we take the curves $\ell_0$ and $\ell$ along with the flopped curve $G'$ as depicted in figure \ref{fig:GlobalThree}.  The intersection numbers of these curves with various divisors are presented in the following table

\begin{center}
\begin{tabular}{|r||r|r|r|r|}
\hline
${}$        & $H$ & $E'$ & $H-D'-E'$ & $D'$ \cr\hline\hline
$\ell_0$    & $1$ & $0$ & $+1$  & $0$ \cr\hline
$\ell$      & $0$ & $-1$ & $0$  & $1$\cr\hline
$G'$    & $0$ & $-1$ & $-1$ & $2$\cr\hline
$\ell-G'$   & $0$ & $0$ & $+1$ & $-1$\cr\hline
\end{tabular}
\end{center}

\noindent The intersections of the divisors with one another are as follows

\begin{center}
\begin{tabular}{|r||c|c|c|}
\hline
        &$H$    & $E'$  &   $H-D'-E'$   \cr\hline\hline
$H$     & $\ell_0   $& $0 $& $\ell_0 - 3l+ 3 G' $    \cr\hline
$E'$        & $0$   &$ - 2\ell+ G'$ & $G'$  \cr\hline
$H-D'-E'$   & $\ell_0 - 3\ell + 3 G'$ & $G'$ & $- 2 \ell_0+ 6 \ell - 5 G' $  \cr\hline
$D'$    & $3 \ell- 3 G'$ &  $2\ell - 2 G' $ & $3 \ell_0 - 9 \ell+ 7 G'$ \cr\hline
\end{tabular}
\end{center}

\noindent
It is useful to distinguish the two $\mathbb{P}^1$'s of $E'$ that are equivalent to $\ell$ inside $\tilde{X}$.  Denoting these by $\ell_1$ and $\ell_2$, we find that
\be\label{NewEll}
{E'}^2 = G' - \ell_1 -\ell_2 \,,\qquad
D'.E' = (\ell_1 - G') + (\ell_2 - G') \,.
\ee

The non-vanishing triple intersection numbers are easily computed from the above data with the following results
\be
\ba
H^3 &= 1 \cr
E^{\prime\,3}&= 1 \cr
D^{\prime\,3}&= -6 \cr
D^{\prime\,2}H &= -3 \cr
D^{\prime\,2}E' &= -2 \,.
\ea
\ee

In the previous section  we listed various divisors and their corresponding holomorphic sections on $X$.
Each of these carries over to a divisor or section after the flop.
We will abuse notation in what follows and continue to use the labels $Z_i,W_j,V_k$ of \eqref{SectionsX} for the corresponding holomorphic sections on $\tilde{X}$.

We use the standard basis for $S_{GUT}=dP_2$ consisting of the hyperplane class, $h$, and the two exceptional curves, $e_1$ and $e_2$
\be
H_2(E',\mathbb{Z}) = \langle h,e_1,e_2\rangle\,.
\ee
From the intersection form
\be
h^2=1\,,\qquad e_i\cdot e_j=-\delta_{ij}\,,
\ee
it is easy to obtain the relation of these classes to $\ell_1$, $\ell_2$, and $G'$,
\be\label{CurveIdent}
\ba
\ell_1 &= h-e_1 \\
\ell_2 &= h-e_2 \\
G' &= h-e_1-e_2\,. \ea \ee

Finally note that $t=-c_1({\cal N}S_{\rm GUT})=-E'^2\vert_{E'}=h$ where we use that
${\cal N}S_{\rm GUT}=E'$ and the fact that the class $2l-G'$ in $\tilde X$ restricts to the class
$h$ in $E'=dP_2.$


\section{Incomplete GUT Multiplets and MSSM $\beta$ Functions}
\label{app:mattercurves}

In this Appendix, we study corrections to the MSSM $\beta$ functions that arise from incomplete GUT multiplets engineered on $\mathbf{10}$ and $\mathbf{5}$ matter curves threaded by nontrivial hypercharge flux.  As discussed in section \ref{subsubsec:unification}, internal hypercharge flux splits the gauge couplings at the GUT scale, $\alpha_i^{-1}(M_{GUT})$, replacing the condition of complete unification with the weaker one \cite{Blumenhagen:2008aw}
\begin{equation}\alpha_1^{-1}(M_{GUT})-\frac{3}{5}\alpha_2^{-1}(M_{GUT})-\frac{2}{5}\alpha_3^{-1}(M_{GUT})=0\,.\label{appblumcond}\end{equation}
Matter fields that comprise incomplete GUT multiplets will disrupt unification in a manner consistent with this provided their contributions, $\delta b_i$, to the MSSM $\beta$ function coefficients, $b_i$, satisfy
\begin{equation}\delta b_1 - \frac{3}{5}\delta b_2 - \frac{2}{5}\delta b_3 = 0\,.\label{appblumenhagenconstraint}\end{equation}
In this Appendix, we verify that this condition is satisfied for any incomplete GUT multiplets that are engineered on $\mathbf{10}$ or $\mathbf{5}$ matter curves threaded by nontrivial hypercharge flux.

\subsection{$\beta$ Functions}

First, let us review some elementary aspects regarding $\beta$ functions in the MSSM.  The coefficients $b_i$ enter into the RG running via
\begin{equation}\frac{d\alpha_i}{dt} = -\frac{b_i}{2\pi}\alpha_i^2\, ,\end{equation}
where we use the normalizations of \cite{Blumenhagen:2008aw}.  For $SU(N)$ gauge theories with fundamental matter, one has
\begin{equation}b_N = \frac{11}{3}N - \frac{1}{3}n_f - \frac{1}{6}n_s\,,\end{equation}
where $n_f$ is the number of left-handed fermions (counting right-handed fermions as left-handed antiparticles) and $n_s$ is the number of complex scalars that couple to gauge bosons.  In a theory with ${\cal{N}}=1$ supersymmetry, there will be gauginos that contribute $-\frac{2}{3}N$ to the running so that the net contribution from gauge degrees of freedom is the usual $\frac{11}{3}N - \frac{2}{3}N = 3N$.  On the other hand, fundamental matter comes from chiral superfields, each of which contains one complex scalar degree of freedom and one left-handed fermion.  As a result, we get the usual formula
\begin{equation}b_N = 3N-\frac{N_f}{2}\, ,\end{equation}
where $N_f$ is the number of chiral superfields in the fundamental of $SU(N)$.  For $U(1)_Y$, there is no contribution from gauge bosons or gauginos.  The contribution from fermions of charge $Y_f$ and scalars of charge $Y_s$ is given by
\begin{equation}-\frac{2}{3}\sum_fY_f^2 - \frac{1}{3}\sum_s Y_s^2\, ,\end{equation}  
so that for a single chiral superfield we get the contribution $-\frac{2}{3}Y^2 - \frac{1}{3}Y^2 = -Y^2$.  Note, however, that the correctly normalized $U(1)_Y$ generator has in addition a factor of $\sqrt{\frac{3}{5}}$.  This  means that the $\beta$ function coefficient for hypercharge is given by
\begin{equation}b_1 = -\frac{3}{5}\sum_{\text{flavors}} Y^2\, .\end{equation}
It is now easy to verify that the matter content of the MSSM gives rise to the usual $\beta$ function coefficients
\begin{equation}b_3=3\, ,\qquad b_2=-1\, ,\qquad b_1=-\frac{33}{5}\, .\end{equation}

\subsection{Spectrum on Matter Curves with Nontrivial $U(1)_Y$ Flux}

Let us now consider how the spectrum on a matter curve with nontrivial $U(1)_Y$ flux affects the $\beta$ function coefficients.

\subsubsection{$\mathbf{5}$ Matter Curves}

We begin with a $\mathbf{5}$ matter curve, $\Sigma_{\mathbf{5}}$, which houses two types of $SU(3)\times SU(2)\times U(1)_Y$ multiplet (and their conjugates), namely
\begin{equation}(\mathbf{3},\mathbf{1})_{-1/3}\oplus (\mathbf{1},\mathbf{2})_{1/2}\, .\end{equation}
In the presence of a bulk flux that engineers $M$ complete $\mathbf{5}$ multiplets and $N$ units of hypercharge flux{\footnote{By hypercharge flux, we mean here the bundle $L_Y^{5/6}$ in \cite{Donagi:2008kj} that is conventionally taken to be ${\cal{O}}(e_1-e_2)$ where $e_1,e_2$ are exceptional classes of the underlying $dP_n$ surface.}}, the net chiralities of these types of multiplets are given by
\begin{equation}\begin{split}n_{(\mathbf{3},\mathbf{1})_{-1/3}}-n_{(\mathbf{\overline{3}},\mathbf{1})_{+1/3}} &= M \\
n_{(\mathbf{1},\mathbf{2})_{+1/2}}-n_{(\mathbf{1},\mathbf{2})_{-1/2}} &= M+N\, .\end{split}\end{equation}
The shift of $\beta$ function coefficients induced by the extra $(\mathbf{3},\mathbf{1})_{-1/3}$'s and $(\mathbf{1},\mathbf{2})_{+1/2}$'s is given by
\begin{equation}\begin{split}
\delta b_3 &= -\frac{M}{2} \\
\delta b_2 &= -\frac{M+N}{2} \\
\delta b_1 &= -\frac{1}{10}\left(5M+3N\right)\, .
\end{split}\end{equation}
Quite nicely, these satisfy \eqref{appblumenhagenconstraint} because
\begin{equation}\delta b_1 - \frac{3}{5}\delta b_2 - \frac{2}{5}\delta b_3 = 0\, .\end{equation}
This result was already obtained in \cite{Blumenhagen:2008aw}, where it was observed that massive fields with the $SU(5)$ quantum numbers of Higgs triplets, which can be engineered by themselves on a $\mathbf{5}$ matter curve with suitable fluxes, can split the gauge couplings at $M_{GUT}$ while retaining the condition \eqref{appblumcond}.

\subsubsection{$\mathbf{10}$ Matter Curves}

We turn now to a $\mathbf{10}$ matter curve, $\Sigma_{\mathbf{10}}$, which houses three types of $SU(3)\times SU(2)\times U(1)_Y$ multiplets (and their conjugates), namely
\begin{equation}(\mathbf{3},\mathbf{2})_{+1/6}\oplus (\mathbf{\overline{3}},\mathbf{1})_{-2/3}\oplus (\mathbf{1},\mathbf{1})_{1}\, .\end{equation}
Here, combining a bulk flux that normally engineers $M$ complete $\mathbf{10}$'s with $N$ units of hypercharge flux leads to the following net chiralities in the spectrum
\begin{equation}\begin{split}
n_{(\mathbf{3},\mathbf{2})_{+1/6}}-n_{(\mathbf{\overline{3}},\mathbf{2})_{-1/6}} &= M \\
n_{(\mathbf{\overline{3}},\mathbf{1})_{-2/3}}-n_{(\mathbf{3},\mathbf{1})_{+2/3}} &= M-N \\
n_{(\mathbf{1},\mathbf{1})_1}-n_{(\mathbf{1},\mathbf{1})_{-1}} &= M+N\, .
\end{split}\end{equation}
In this case, the shift of the $\beta$ coefficients induced by the extra $(\mathbf{3},\mathbf{2})_{+1/6}$'s, $(\mathbf{\overline{3}},\mathbf{1})_{-2/3}$'s, and $(\mathbf{1},\mathbf{1})_{+1}$'s is given by
\begin{equation}\begin{split}
\delta b_3 &= -\frac{1}{2}\left(3M-N\right) \\
\delta b_2 &= -\frac{3}{2}M \\
\delta b_1 &= -\frac{1}{10}\left(15M-2N\right)\, .
\end{split}\end{equation}
Quite nicely, these also satisfy \eqref{appblumenhagenconstraint} because
\begin{equation}\delta b_1 - \frac{3}{5}\delta b_2 - \frac{2}{5}\delta b_3 = 0\, .\end{equation}

\newpage

\bibliographystyle{JHEP}
\renewcommand{\refname}{Bibliography}
\addcontentsline{toc}{section}{Bibliography}

\providecommand{\href}[2]{#2}\begingroup\raggedright\endgroup


\end{document}